\documentclass[preprint,aps,amssymb,showpacs,superscriptaddress,nofootinbib]{revtex4}
\usepackage{graphicx}

\newtheorem{prop}{Proposition}
\newtheorem{lemma}{Lemma}

\newcommand{\Case}[2]{{\textstyle \frac{#1}{#2}}}
\newcommand{\lP}{\ell_{\mathrm P}}

\newcommand{\md}{{\mathrm{d}}}
\newcommand{\tr}{\mathop{\mathrm{tr}}}
\newcommand{\sgn}{\mathop{\mathrm{sgn}}}

\begin{document}
\preprint{IMSc/2004/04/16}
\preprint{AEI--2004--028}

\title{The Bianchi IX model in Loop Quantum Cosmology}

\author{Martin Bojowald}
\email{mabo@aei.mpg.de}
\affiliation{Max-Planck-Institut f\"ur Gravitationsphysik,
Albert-Einstein-Institut,\\
Am M\"uhlenberg 1, D-14476 Golm, Germany}
\author{Ghanashyam Date}
\email{shyam@imsc.res.in}
\affiliation{The Institute of Mathematical Sciences\\
CIT Campus, Chennai-600 113, INDIA.}
\author{Golam Mortuza Hossain}
\email{golam@imsc.res.in}
\affiliation{The Institute of Mathematical Sciences\\
CIT Campus, Chennai-600 113, INDIA.}

\begin{abstract}
The Bianchi IX model has been used often to investigate the structure
close to singularities of general relativity. Its classical chaos is
expected to have, via the BKL scenario, implications even for the
approach to general inhomogeneous singularities. Thus, it is a popular
model to test consequences of modifications to general relativity
suggested by quantum theories of gravity. This paper presents a
detailed proof that modifications coming from loop quantum gravity
lead to a non-chaotic effective behavior. The way this is realized, 
independently of quantization ambiguities, suggests a new look at initial 
and final singularities.
\end{abstract}

\pacs{0460P, 0460K, 9880H}

\maketitle

\section{Introduction}

It is well known that according to the singularity theorems
\cite{HawkingEllis}, the space-time describing the backward evolution
of an expanding universe is necessarily singular in the sense of
geodesic incompleteness. The generality of singularity theorems also
prevents them from giving any information about the {\em nature} of
such a singularity for example in terms of curvature invariants. By
contrast, the BKL approach to the issue of singularities asks
\cite{BKL}: In a neighborhood of a presumed singularity, is there {\em
a general} solution of the Einstein equation such that at least some
curvature invariants diverge? In this formulation of the question, one
also obtains information about the possible nature of the presumed
singularity described in terms of the {\em approach to the
singularity}.  The conclusion of the long and detailed analysis is
summarized in the BKL scenario: Generically, as the singularity is
approached, the spatial geometry can be viewed as a collection of
small patches each of which evolves essentially independently of the
others according to the Bianchi IX evolution (for recent numerical
evidence see \cite{Garfinkle}). The Bianchi IX evolution towards its
singularity is described by an infinite succession of Kasner
evolutions (two directions contracting while third one expanding)
punctuated by permutations of the expanding/contracting directions as
well as possible rotations of the three directions themselves. The
qualitative analysis of these permutations and rotations
(`oscillations') is done very conveniently in terms of a billiard ball
bouncing off moving walls and has been analyzed for a possible {\em
chaotic} behavior \cite{Billiards}.

However, it has been well recognized that the conclusion of ad
infinitum oscillations of the BKL singularity, a consequence of
unbounded growth of the spatial curvature, cannot be trusted very
close to the singularity where the classical Einstein equations
themselves are expected to break down. Presumably the equations will
be superceeded by some quantum extensions. In the absence of any
specific and detailed enough quantum theory the questions of the fate
of the classical singularity vis a vis the behavior of space-time near
such a region, could not have been addressed. The focus therefore has
been to include qualitatively expected quantum modification. For
example, in the Kaluza-Klein picture and more recently the stringy
picture, the minimal expected modification is the matter content --
notably the dilaton and p-form fields. Higher derivative corrections
in the effective action are also expected. Another qualitative
implication of string theory namely the brane world scenarios is also
a candidate to study implications of modified Einstein
equations. All of these have been explored with varying conclusions
regarding the BKL behavior, e.g.\
\cite{StringChaos,BraneNonChaos}. However, they {\em still
contain the singularity} and hence all these modified equations must
break down near the singularity raising questions about the validity
of the conclusion.

In recent times this situation has changed, particularly in the
cosmological context. Within the loop quantum gravity approach
\cite{Revs}, we do have the detailed framework of
LQC \cite{LoopCosRev} to deduce the nature of quantum effects.  Due to
the underlying quantum geometry \cite{AreaVol,Area,Vol2,cosmoII}, the
internal time (volume) evolution becomes a difference equation
\cite{cosmoIII,cosmoIV} and one also has a quantization of the inverse
scale factor with a bounded spectrum
\cite{InvScale}. Thus the classically indicated singularity is {\em
removed} i.e.\ the evolution extends to `negative' (internal) time and
the spatial curvatures remain finite
\cite{Sing,IsoCosmo,HomCosmo,Spin}. The equations also admit a
continuum approximation thereby recovering the classical continuum
geometry for larger volumes \cite{SemiClass,FundamentalDisc}.

It is now meaningful to ask if and how the BKL behavior is modified.
There is in fact a hint of what may be expected. For larger volumes,
one can trust the classical picture. Then, in the general
inhomogeneous cosmological context, one can approximate smaller
patches of the spatial geometry by the homogeneous Bianchi IX
models. As the volume is decreased, the patches have to be made
smaller to sustain the approximation. {\em If} the BKL behavior were
to continue for these individual patches, then the fragmentation into
smaller patches must continue ad infinitum. But the underlying
discrete structure cannot support such infinite
fragmentation. Therefore, the quantum geometry which is responsible
for singularity removal, must also ensure that Bianchi IX behavior may
at the most have finitely many oscillations.  This is in fact a self
consistency requirement on the procedures of LQC. As reported in
\cite{NonChaos}, the {\em self consistency} indeed holds and we
elaborate the proof in this paper.

The paper is organized as follows.

In section II, we recall the relevant details of the classical and the loop 
quantized description of the vacuum Bianchi IX model. We describe the 
kinematical set-up, the Hamiltonian constraint and the definition of inverse 
triad components including a class of quantization ambiguities. 

In section III, we recall the continuum approximation leading to the
Wheeler--DeWitt differential equation. We describe how the {\em effective 
continuum Hamiltonian} is obtained and how the most important {\em 
non-perturbative} corrections from the underlying quantum equations are
incorporated. We also comment on the validity of the effective
description. 

In section IV, we display the effective Hamiltonian for the Bianchi IX
model and specify the dynamics with respect to an {\em internal time}
obtained by solving the effective Hamiltonian constraint.

In section V, we analyze the potential term in the Hamiltonian for
internal dynamics. This section is divided into four subsections. In
subsection A, we describe the `walls' (positive potential) of the
potential. Unlike the classical one, this modified potential has a
non-factorizable dependence on internal time. The walls therefore not
only move but also change their extent and height, eventually
disappearing completely. Subsection B contains a demonstration of the
stability of the Kasner solution when the effective potential is turned
on. In subsection C, we relate the Kasner solution in terms of the
internal time to the more common description in terms of the scale factor 
evolution. In the last subsection, we specify the domain of validity of
the effective description in quantitative terms.

In section VI, we extrapolate beyond the domain of validity mainly to
contrast the usual oscillatory behavior under the influence of the
classical potential. The (small volume) asymptotic analysis of this section 
uses qualitative arguments and proceeds in two main steps. In the first
step, we rule out motion in the anisotropy plane being confined to a bounded 
region. In the second step we show that the asymptotic trajectory must
approach a Kasner trajectory exponentially. 

In section VII, we summarize our results. We also point out the insensitivity
of the results to quantization ambiguities and discuss possible
implications  of the modified dynamics as to the structure of initial and 
final singularities.

\section{The Bianchi IX model}

Bianchi models are a subclass of spatially homogeneous models such that 
the symmetry group contains a subgroup with {\em simply transitive} action on 
the spatial manifold.  The simply transitive subgroups are classified in 
terms of three integers, $n^I$, parameterizing the structure constants as
\[
C^I_{JK}=\epsilon_{(I)JK}n^{I} \ .
\] 

Bianchi IX class of models have $n^I = 1,  I = 1, 2, 3$. Clearly, 
$C^I_{JI}=0$ implying that Bianchi IX is a class A model admitting a canonical
formulation.

Given the subgroup acting simply transitively, one introduces the
corresponding left-invariant 1-forms $\omega^I = \omega^I_a \md x^a$
and their dual, density weighted vector fields $X_I = X^a_I
\partial_a$, on the spatial manifold $\Sigma$.
The 1-forms satisfy the Maurer-Cartan equations,
\[
 \md\omega^I=-\Case{1}{2}C^I_{JK}\omega^J\wedge\omega^K\,.
\]

In the metric description, the metrics of Bianchi IX models have the
form,
\begin{equation}
ds^2 ~=~ dt^2 - \gamma_{IJ}(t) \omega^I \omega^J,
\end{equation}
while in the connection formulation, the geometry is described in
terms of the invariant connections and densitized triads given by,
\[
 A^i_a ~ = ~ \phi^i_I (t) \omega^I_a ~~,~~ E^a_i ~ = ~ p^I_i (t)  X^a_I\, 
\]
with spatially constant and canonically conjugate $\phi^i_I$, $p^I_i$. 

Note that while the symmetry group `fixes' the invariant 1-forms only 
up to orthogonal transformations which preserve the structure constants of 
Bianchi IX, this freedom can be used only at {\em one} instant of
time during an evolution. The $t$ dependence resides in the symmetric 
$\gamma_{IJ}$ in the metric formulation and in $\phi^i_I, p^I_i$ in the 
connection formulation. 

In the vacuum case, the Einstein equations $R_{0a} = 0$, imply that the
matrix $\gamma_{IJ}$ can be taken to be {\em diagonal} so that only
three degrees of freedom remain in the metric formulation. In the presence
of matter, the metric is in general non-diagonal. However, as argued by
BKL, near the singularity, the non-diagonal terms do {\em not} affect
either the nature of Kasner epochs (time intervals during which scales of
spatial distances along three specific directions change as in the case
of the Kasner solution), or the `law of replacement of Kasner
exponents'. Instead, these terms lead to rotation of the Kasner axes
themselves during a change over from one Kasner epoch to another. Thus,
to study the `oscillatory behavior' implied by succession of Kasner
epochs with changes in the exponents, it is sufficient to restrict to
the sub-class of {\em diagonal} Bianchi IX, and anyway for the vacuum
case this restriction is automatically implied. One now has only three
degrees of freedom in the metric formulation.

In the connection formulation the situation is analogous. To
begin with one has nine degrees of freedom, $\phi^i_I$ but with a
three parameter gauge freedom of SU(2) rotations of the index $i$. A
further freedom, depending upon the parameters $n^I$, would have been
available if the left-invariant 1-forms were not thought of as a
background structure in parameterizing the invariant connections. To
get the same three degrees of freedom as in the diagonal metric
formulation, one then restricts the $\phi^i_I$ to a ``diagonal form",
$\phi^i_I := c_{(I)} \Lambda^i_I$ and correspondingly $p^I_i := p^I
\Lambda^{(I)}_i$ where the SO(3)-matrix $\Lambda$ includes gauge
degrees of freedom (see \cite{HomCosmo} for details). Such a
restriction specifies the {\em diagonal models} in the connection
formulation. This is {\it not} a symmetry reduction in the same sense
as restriction to isotropic models is. Nevertheless, these restricted
models can be analyzed by following procedures similar to those in the
context of symmetry reductions \cite{SymmRed}. As in reductions to
isotropic models, this leads to significant simplifications in such
diagonalized models because the gauge parameters in $\Lambda$ and
(almost; see below) gauge invariant parameters, $c_I$, are neatly
separated \cite{HomCosmo}.

\subsection{Classical Framework}
\label{ModClass}

The basic variables for diagonal Bianchi Class A models are specified
via
\begin{equation}
 A_a^i=c_{(I)}\Lambda_I^i\omega_a^I
\end{equation}
in terms of the `gauge invariant' coefficients $c_I$ with the pure
gauge degrees of freedom contained in the $SO(3)$-matrix
$\Lambda$. The components $c_I$ are not completely gauge invariant
but subject to residual discrete gauge transformations which change
the sign of two of the three components simultaneously. A diagonal
densitized triad has the form
\begin{equation}
 E_i^a=p^{(I)}\Lambda^I_iX_I^a
\end{equation}
where $X_I$ are left-invariant densitized vector fields dual to
$\omega^I$. Being an SO(3)-matrix, $\Lambda$ satisfies
\[
\Lambda^i_I \Lambda^J_i ~ = ~ \delta_I^J ~~ , ~~ 
\epsilon_{ijk} \Lambda^i_I \Lambda^j_J \Lambda^k_K ~ = ~ \epsilon_{IJK}
\]

The triad components $p^I$ are subject to the same residual gauge
transformations as the connection components $c_I$ to which they are
conjugate with basic Poisson bracket
\begin{equation}
 \{c_I,p^J\}=\gamma\kappa\delta_I^J
\end{equation}
where $\gamma$ is the Barbero--Immirzi parameter and $\kappa=8\pi G$
the gravitational constant. We will use the value
$\gamma=\frac{\log(2)}{\pi \sqrt{3}}\approx0.13$ fixed by the black
hole entropy calculations \cite{ABCK:LoopEntro,IHEntro}.

A diagonal co-triad has the form $e_a^i=a_{(I)}\Lambda_I^i\omega_a^I$
with
\begin{equation}
 p^1=|a_2a_3|\sgn(a_1)\quad,\quad p^2=|a_1a_3|\sgn(a_2) \quad,\quad
 p^3=|a_1a_2|\sgn(a_3)\,.
\end{equation}
Note that the components $p^I$ as well as $a_I$ can take negative
values, but only the overall sign $\sgn(p^1p^2p^3)$ (i.e., the
orientation) and the absolute values $|p^I|$ are gauge invariant. With
this form of the co-triad, we obtain, in fact, a diagonal metric
\[
 \md s^2=e_I^ie_J^i\omega^I\omega^J=\sum_Ia_I^2(\omega^I)^2\,.
\]

The extrinsic curvature and the spin connection components likewise
have a diagonal form: $k_a^i := K_{(I)}\Lambda_I^i\omega_a^I$ and
$\Gamma_a^i := \Gamma_{(I)}\Lambda_I^i\omega_a^I$ with $K_I=-
\Case{1}{2}\dot{a}_I$ which appear in the relation
$c_I=\Gamma_I-\gamma K_I$ defining the Ashtekar connection components
$c_I$ in terms of the spin connection components $\Gamma_I$ and the
extrinsic curvature. The spin connection components for Bianchi IX are
given by \cite{HomCosmo},
\begin{eqnarray} 
 \Gamma_I &=& \Case{1}{2}\left(\frac{a_J}{a_K}+ \frac{a_K}{a_J}-
   \frac{a_I^2}{a_Ja_K}\right)\\
  &=& \Case{1}{2}\left(\frac{p^K}{p^J}+ \frac{p^J}{p^K}-
   \frac{p^Jp^K}{(p^I)^2}\right) \mbox{ for $(I,J,K)$ an even
   permutation of $(1,2,3)$.} \label{SpinConn}
\end{eqnarray}
Note that in contrast to the full theory, the spin connection is a
covariant object within a homogeneous model since coordinate
transformations have to respect the symmetry. Consequently, if
non-zero, it cannot be made small by choosing appropriate local
coordinates. 

The Hamiltonian constraint \cite{HomCosmo} is given by,
\begin{eqnarray} \label{H}
 H &=& 2\kappa^{-1}\left\{\left[(c_2\Gamma_3+c_3\Gamma_2-\Gamma_2\Gamma_3)
     (1+\gamma^{-2})- c_1-\gamma^{-2}c_2c_3\right]a_1\right.\nonumber\\
  &&+\left[(c_1\Gamma_3+c_3\Gamma_1-\Gamma_1\Gamma_3)
     (1+\gamma^{-2})- c_2-\gamma^{-2}c_1c_3\right]a_2\nonumber\\
  &&\left.+\left[(c_1\Gamma_2+c_2\Gamma_1-\Gamma_1\Gamma_2)
     (1+\gamma^{-2})- c_3-\gamma^{-2}c_1c_2\right]a_3\right\}\\
 &=& 2\kappa^{-1}\left[(\Gamma_2\Gamma_3-\Gamma_1)a_1+
   (\Gamma_1\Gamma_3-\Gamma_2)a_2
   +(\Gamma_1\Gamma_2-\Gamma_3)a_3\right.\nonumber\\
  &&\left. -\Case{1}{4}(a_1\dot{a}_2\dot{a}_3+ a_2\dot{a}_1\dot{a}_3+
   a_3\dot{a}_1\dot{a}_2)\right]\,. \label{HH}
\end{eqnarray}
In order to derive the classical field equations (for which we
consider only positive $a_I$ and $p^I$) it is advantageous to
transform to new canonical variables
\begin{eqnarray} \label{CanonicalVar}
\pi_I &:=& 2K_Ip^{(I)}= - \dot{a}_Ia_I^{-1}a_1a_2a_3= - (\log a_I)'
\quad\mbox{ and} \nonumber \\ 
q^I &:=& \Case{1}{2}\log p^I
\hspace{2cm}\mbox{such that}\hspace{2cm}\{q^I, \pi_J\} = \kappa
\delta_I^J 
\end{eqnarray}
where the prime denotes a derivative with respect to a new time
coordinate $\tau$ related to $t$ by $\md t=a_1a_2a_3\md\tau$
(corresponding to the lapse function $N=a_1a_2a_3$). With these new
variables we have $\{\pi_I,p^J\}= - 2\kappa p^{(I)}\delta_I^J$ and
\begin{eqnarray}
 \kappa NH &=& \kappa a_1a_2a_3H=
 2\left[p^1p^2(\Gamma_1\Gamma_2-\Gamma_3)+
   p^1p^3(\Gamma_1\Gamma_3-\Gamma_2)+
   p^2p^3(\Gamma_2\Gamma_3-\Gamma_1)\right.\nonumber\\
   \label{ClassHam}
 && \left. - \Case{1}{4}(\pi_1\pi_2+\pi_1\pi_3+\pi_2\pi_3)\right]\\
 &=& \Case{1}{2}\left[\left(\frac{p^2p^3}{p^1}\right)^2+
   \left(\frac{p^1p^3}{p^2}\right)^2+
   \left(\frac{p^1p^2}{p^3}\right)^2\right]\nonumber\\
 &&- (p^3)^2-
 (p^2)^2- (p^1)^2
 -\Case{1}{2}(\pi_1\pi_2+\pi_1\pi_3+\pi_2\pi_3)\\
 &=& \Case{1}{2}\left(a_1^4+ a_2^4+ a_3^4\right)-
 a_1^2a_2^2- a_1^2a_3^2- a_2^2a_3^2
 -\Case{1}{2}(\pi_1\pi_2+\pi_1\pi_3+\pi_2\pi_3)\,.
\end{eqnarray}
Now one can easily derive the equations of motion,
\[
 (\log a_I)''= - \pi_I' \approx - \{\pi_I,a_1a_2a_3H\}
\]
which yield
\begin{eqnarray} \label{motion}
 \Case{1}{2}(\log a_1)'' &=& (a_2^2-a_3^2)^2-a_1^4\nonumber\\
 \Case{1}{2}(\log a_2)'' &=& (a_1^2-a_3^2)^2-a_2^4\\
 \Case{1}{2}(\log a_3)'' &=& (a_1^2-a_2^2)^2-a_3^4\nonumber
\end{eqnarray}
or, using $(q^1)''= - \Case{1}{2}(\pi_2'+\pi_3')$,
\[
 (q^I)''= -4a_I^2p^I\Gamma_I\,. 	
\]

For the Bianchi I model we would have the right hand sides of equation
(\ref{motion}) being zero so that $\log
a_I=\alpha_I(\tau-\tau_{0,I})$. This implies the Kasner behavior
$a_I\propto t^{\alpha_I}$ where $t=e^\tau$ and the constraint
(eq.(\ref{ClassHam}) with $\Gamma_I = 0$) requires
$0=\alpha_1\alpha_2+\alpha_1\alpha_3+\alpha_2\alpha_3=
\Case{1}{2}((\alpha_1+\alpha_2+\alpha_3)^2-
\alpha_1^2-\alpha_2^2-\alpha_3^2)$. The coefficients $\alpha_I$ can be
rescaled by choosing a different $t(\tau)$ which can be fixed by
requiring the conventional parameterization
$\alpha_1+\alpha_2+\alpha_3=1= \alpha_1^2+\alpha_2^2+\alpha_3^2$. As
usual, these equations can be solved only if one coefficient, say
$\alpha_1$, is negative and the other two are
positive. Correspondingly, one direction, the first, contracts whereas
the other two expand toward larger time. When we approach the
classical singularity at $t=0$, space shrinks only in two directions
while the third one expands unboundedly. The total volume however
continues to approach zero according to $a_1a_2a_3\propto
t^{\alpha_1+\alpha_2+\alpha_3}=t$. In \cite{HomCosmo} it has been
shown that the Kasner singularity disappears when the model is
quantized along the lines of loop quantum cosmology. 

For Bianchi IX the evolution of the three triad components can be
described as motion in a non-trivial potential given by
\begin{eqnarray} \label{potential}
 W(p^1,p^2,p^3) &=& 2\left\{p^1p^2(\Gamma_1\Gamma_2-\Gamma_3)+
 p^1p^3(\Gamma_1\Gamma_3-\Gamma_2)+
 p^2p^3(\Gamma_2\Gamma_3-\Gamma_1)\right\}
\end{eqnarray}
which has infinite walls at small $p^I$ owing to the divergence of the
spin connection components. The evolution can then be described approximately
as a succession of Kasner epochs with intermediate reflections at the 
potential \cite{Mixmaster}. For the Bianchi IX model the reflections never 
stop and the classical evolution is believed to be chaotic \cite{Chaos}.

\subsection{Loop Quantization}
\label{s:Loop}

Diagonal homogeneous loop quantum cosmology \cite{HomCosmo} is first
formulated in the connection representation where an orthonormal basis
is given by the $\hat{p}^I$-eigenstates
\begin{equation}\label{n}
 |m_1,m_2,m_3\rangle:= |m_1\rangle\otimes |m_2\rangle\otimes
 |m_3\rangle
\end{equation}
with
\begin{equation} \label{cm}
 \langle c|m\rangle=\frac{\exp(\Case{1}{2}imc)}{\sqrt{2}\sin(\Case{1}{2}c)}\,.
\end{equation}
Here we are already restricting ourselves to a separable subspace of
the kinematical Hilbert space, which is non-separable \cite{Bohr}. All
the information about the classical singularity is already contained
in wave functions restricted to this subspace such that it suffices
for our purposes.

The eigenvalues of the triad operators can be read off
from
\begin{equation}
 \hat{p}^I|m_1,m_2,m_3\rangle= \Case{1}{2}\gamma\lP^2m_I
 |m_1,m_2,m_3\rangle\,.
\end{equation}
Using the basic operators $\hat{p}^I$ one can define the volume
operator $\hat{V}=\sqrt{|\hat{p}^1\hat{p}^2\hat{p}^3|}$ which will be
used later. Its eigenstates are also $|m_1,m_2,m_3\rangle$ with
eigenvalues
\begin{equation} \label{V}
 V(m_1,m_2,m_3)=(\Case{1}{2}\gamma\lP^2)^{\frac{3}{2}}
 \sqrt{|m_1m_2m_3|}\,.
\end{equation}

A kinematical state $|s\rangle$ is described in the triad representation
by coefficients $s_{m_1, m_2, m_3}$ defined via
\begin{equation}\label{TriadRep}
 |s\rangle=\sum_{m_1, m_2, m_3}s_{m_1, m_2, m_3}|m_1,m_2,m_3\rangle\,.
\end{equation}
For a state to be gauge invariant under the residual gauge
transformations changing the sign of two $p$-components
simultaneously, the coefficients $s_{m_1,m_2,m_3}$ have to satisfy
\begin{equation}
 s_{m_1,m_2,m_3}=s_{-m_1,-m_2,m_3}=s_{m_1,-m_2,-m_3}=
 s_{-m_1,m_2,-m_3}\,.
\end{equation}
These states are left invariant by the gauge invariant triad operators
$|\hat{p}^I|$ and the orientation operator
$\sgn(\hat{p}^1\hat{p}^2\hat{p}^3)$. In calculations it is often
easier to work with non-gauge invariant states in intermediate steps
and project to gauge invariant ones in the end.

Together with the basic derivative operators $\hat{p}^I$ we need
multiplication operators which usually arise from (point) holonomies
$h_I=\exp(c_{(I)}\Lambda_I^i\tau_i)=
\cos(\Case{1}{2}c_I)+2\Lambda_I^i\tau_i \sin(\Case{1}{2}c_I)$ with
action
\begin{eqnarray}
 \cos(\Case{1}{2}c_1) |m_1,m_2,m_3\rangle &=&
 \Case{1}{2}(|m_1+1,m_2,m_3\rangle+|m_1-1,m_2,m_3\rangle) \label{cos}\\ 
 \sin(\Case{1}{2}c_1) |m_1,m_2,m_3\rangle &=&
 -\Case{1}{2}i(|m_1+1,m_2,m_3\rangle- |m_1-1,m_2,m_3\rangle) \label{sin}
\end{eqnarray}
and correspondingly for $c_2$ and $c_3$.

\subsubsection{Inverse triad operators}

{}From the basic operators we can build more complicated ones. We will
later need a quantization of the spin connection which is a composite
operator containing several triad operators. In particular, it also
contains inverse powers of triad components which classically diverge
at the singularity. Since the triad operators have a discrete spectrum
containing zero, they do not have an inverse. However, general methods
of quantum geometry and loop quantum cosmology \cite{QSDV,InvScale}
imply that there exist well-defined operators quantizing inverse triad
components. To obtain these operators one makes use of a classical
reformulation, e.g.  
\begin{equation}
|p^1|^{-1}=(l \gamma\kappa)^{\frac{1}{l -1}}(\{c_1,|p^1|^l
\})^{\frac{1}{1 - l}}
= (l \gamma\kappa)^{\frac{1}{l -1}} 
\left( 
\frac{\tr_j\Lambda_1^i\tau_ih_1\{h_1^{-1},|p^1|^l \}} 
{\frac{1}{3}j(j + 1)(2j + 1)}\right)
^ {\frac{1}{1 -l}}
\end{equation}
which can then be quantized to
\begin{equation}\label{InvOpr}
\widehat{|p^1|^{-1}_{j,l}}= {\mathrm i}^{\frac{1}{l -1}}
\left[\gamma\lP^{2} l j(j+1)(2j+1)/3\right]^{\frac{1}{l -1}}
\left(\tr\nolimits_j\Lambda_1^i\tau_i
h_1[h_1^{-1},|\hat{p}^1|^l ] \right)^{\frac{1}{1 -l}} \,.
\end{equation}

Here we have indicated that there are quantization ambiguities
\cite{Ambig,ICGC} when one quantizes composite operators. The two
most relevant for this paper are indicated by the subscript
$(j,l)$. The half integer $j$ appearing as a subscript of the trace
corresponds to the choice of representation while writing holonomies
as multiplicative operators. The real valued $l \in (0, 1)$, labels
the various classically equivalent ways of writing $|p|^{-1}$ as a
Poisson bracket of positive powers of $|p|$.

This operator acts as
\begin{eqnarray} \label{InvOprAct}
\widehat{|p^I|_{j,l}^{-1}}|m_1, m_2, m_3\rangle & := & (2 l)^{\frac{1}{l
-1}}\left(\Case{1}{2}\gamma \ell^2_p\right)^{ -1}
{\cal{N}}_j^{\frac{1}{1-l}} f_{j,l}( m_I) |m_1, m_2, m_3\rangle
~~~\mbox{ where, } \\ 
{\cal{N}}_j & := & \frac{j (j + 1) ( 2 j + 1
)}{3} ~~~ \mbox{and}\nonumber \\ 
f_{j,l}(m_I) & := & \left\{
\sum_{k = - j}^{j} k ( | m_I + 2 k | )^{l} \right\}^{\frac{1}{1 -l}} \, .
\nonumber 
\end{eqnarray}
The discrete values $f_{j,l}(m)$ {\em decrease toward lower values\/}
for $m<2j$ \cite{Ambig}.  Thus, one can see that the classical
divergence of the inverse of $|p^I|$ at vanishing $p^I$ is explicitly
absent in the quantized operator. This will be seen to have further
consequences for the approach to the classical singularity.
Furthermore, the state $|m_1,m_2,m_3\rangle$ with $m_I=0$ (on which
the classical inverse triad would diverge) is actually annihilated by
$\widehat{|p^I|_{j,l}^{-1}}$ due to $f_{j,l}(0)=0$ independently of
the ambiguities $j$ and $l$ in their allowed ranges.\footnote{The usual
case with eigenvalues $m_I^{-1}$, which corresponds to a discrete
version of the Wheeler--DeWitt operator, could be obtained formally from
(\ref{InvOprAct}) with the values $l=2$, $j=1/2$ which are not allowed
here, and which would not result in an operator annihilating the
states corresponding to the classical singularity.}  This allows us to
define the inverse triad operator (not just its absolute value) by
$\widehat{(p^I)_{j,l}^{-1}} := \sgn(\hat{p}^I)
\widehat{|p^I|_{j,l}^{-1}}$ without ambiguity in defining the sign of
zero.  These inverse triad operators will be used later to find
well-defined operators quantizing the spin connection. Their
eigenvalues are
\begin{equation} \label{InvTriad}
 \widehat{(p^I)_{j,l}^{-1}}|m_1, m_2, m_3\rangle = (2 l)^{\frac{1}{l
 -1}} (\Case{1}{2}\gamma \ell^2_p)^{-1} {\cal{N}}_j^{\frac{1}{1-l}}
 \sgn(m_I) f_{j,l}( m_I) |m_1, m_2, m_3\rangle\,.
\end{equation}
It turns out to be convenient to define another function $F_{j,l}(m)$ 
as: 
\begin{equation}
F_{j,l}(m_I) ~ := ~ 2j (2 l {\cal{N}}_j)^{\frac{1}{l - 1}}
f_{j,l}(m_I) \ , 
\end{equation}
so that the eigenvalues are obtained as,
\begin{equation}
\widehat{(p^I)_{j,l}^{-1}}|m_1,m_2,m_3\rangle ~ = ~ 
\left( \frac{1}{2} \gamma \ell_{\text p}^2 \right)^{-1} (2 j)^{-1} 
\sgn(m_I)\ F_{j,l}(m_I) ~ |m_1,m_2,m_3\rangle \ .
\end{equation}
 
Then, for large $j$, using also ${\cal N}_j \sim \Case{2j^3}{3}$,
$F_{j,l}(m_I)=F_l(\mu_I)$ actually turns out to be a function of $\mu_I :=
\Case{m_I}{2 j}$ and with {\em no} dependence on $j$. Explicitly
\cite{Ambig,ICGC},
\begin{eqnarray}
F_l(q) & := &\left( 2l(l+1)(l+2)/3 \right)^{\frac{1}{l - 1}}
\left[(l+1) \left\{ (q + 1)^{l+2} 
- |q - 1|^{l+2} \right\} \right. \nonumber \\
& & \mbox{\hspace{2.0cm}} \left. - (l+2) q \left\{ (q + 1)^{l+1} 
 - \sgn(q - 1) |q - 1|^{l+1} \right\}\right]^{\frac{1}{1-l}}
\label{Fdef}\\
& \longrightarrow & ~~~~q^{-1}  \mbox{\hspace{3.2cm}} (q \gg 1) \nonumber \\
& \longrightarrow & \left[ \frac{3 q}{l+1}\right]^{\frac{1}{1-l}}
\mbox{\hspace{2.0cm}} (0 < q \ll 1) \ , \mbox{\hspace{1.0cm} and}
\label{Fasym} \\
f_{j,l}(m_I) & \approx & \left( 2^{l+1} l/3 \right)^{\frac{1}{1 -l}} 
j^{\frac{2 + l}{1 - l}} \  F_l\left(\mu_I := \Case{m_I}{2j}\right) ~ + ~
o(j^{\frac{1 + 2l}{1 - l}}) \label{Contf} \ .
\end{eqnarray}

There is no suffix $j$ on $F_l$ since it is manifestly independent of $j$. The 
limiting forms show that  for $\mu_I \gg 1$ one has the classical behavior 
for the inverse triad components and the {\em quantum modifications} are 
manifest for $\mu_I < 1$ (or for $m_I < 2j $).

The inverse triad operators are used in defining the spin connection
operators given in eq.(\ref{SpinConn}) (specialized to Bianchi IX) and
the corresponding potential (\ref{potential}). These are diagonal in the
triad representation with eigenvalues given by, 
\begin{eqnarray}
\Gamma_I(\vec{\mu}) & = & \frac{1}{2}\ \left\{
\mu_K \text{sgn}(\mu_J) F_l(\mu_J) + 
\mu_J \text{sgn}(\mu_K) F_l(\mu_K) - 
\mu_J \mu_K F^2_l(\mu_I) \ \right\} \label{SpinCoeff}\\
\frac{W_{j,l}(\vec{\mu})}{\left(\frac{1}{2} \gamma \ell_{\text p}^2 
\right)^2} & = & 8 j^2 \left[ \mu_I \mu_J (\Gamma_I \Gamma_J - \Gamma_K)
+ \text{cyclic} \right] \label{ModPotl}
\end{eqnarray}
Note that the $\Gamma_I$, as functions of $\mu_I$, are both {\em
dimensionless} and {\em independent} of $j$, while the potential
explicitly scales as $j^2$. For $\mu_I \gg 1$, the $\Gamma_I$ become
{\em homogeneous functions of degree zero} of their arguments, an
observation which will be useful while discussing the effective
Hamiltonian. To avoid proliferation of labels, we have suppressed the
ambiguity label $l$ on the spin connection but have displayed both $j,
l$ on the effective potential.
\subsubsection{The Hamiltonian constraint}
The quantization of the Hamiltonian constraint is somewhat more
complicated in the presence of a non-zero spin connection. The details
can be found in \cite{Spin}. Here we summarize only the salient points.

The key difference between the full theory and the homogeneous models
with non-zero spatial curvature is that the spin connection is a {\em
tensor} due to the restriction on the coordinate transformations
preserving the homogeneity. Consequently, it cannot be made to vanish
unlike in the full theory. There is thus no guidance from the full
theory regarding the incorporation of the spin connection. This loss of
guidance is compensated by invoking two {\em criteria of admissibility} of
quantization of the Hamiltonian\cite{FundamentalDisc}. 

Recall that the underlying discrete nature of loop quantum gravity
implies a {\em difference equation form} for imposition of the
Hamiltonian constraint.  The order of this equation determines the
number of independent solutions. Among these, there must be those
which are slowly varying over the discreteness scale which should
correspond to the familiar continuum, classical geometry. In essence,
the two criteria of admissibility stipulate that the difference
equation implied by a quantization of the Hamiltonian, must admit a
{\em continuum approximation} which in addition must be (locally) {\em
stable} in a certain well defined sense.  In \cite{Spin}, a
quantization of the Hamiltonian constraint for all
diagonal homogeneous models was proposed and shown to satisfy the
admissibility criteria.

The quantized Hamiltonian operator is given explicitly as:
\begin{eqnarray}
\hat{H}  & = & 4 i ( \gamma \ell_p^2 \kappa )^{-1} 
\sum_{IJK} \epsilon^{IJK} \tr \left[ ~ \left\{ 
\gamma^{-2} \hat{h}_I(A, \Gamma) \hat{h}_J(A, \Gamma) 
(\hat{h}_I(A, \Gamma))^{-1}(\hat{h}_J(A, \Gamma))^{-1} \right.
\right. \nonumber \\
& & \hspace{4.5cm}
\left. \left. - 2 ( \hat{\Gamma}_I \hat{\Gamma}_J - n^L \hat{\Gamma}_L )
\Lambda_I \Lambda_J \right\} \left\{ h_K(A)
\left[ h_K^{-1}(A) , \hat{V} \right] \right\}~ \right] \label{Hop}
\end{eqnarray}
where
\begin{eqnarray}
\hat{h}_I(A, \Gamma) & := & e^{c_{(I)} \Lambda_I^i \tau_i} ~
e^{- \hat{\Gamma}_{(I)} \Lambda_I^i \tau_i} \label{HolonomyOpr}\\
\hat{\Gamma}_I & := & \frac{1}{2} \left[ 
\hat{p}^J n^K \widehat{(p^K)^{-1}} + 
\hat{p}^K n^J \widehat{(p^J)^{-1}} - 
\hat{p}^J \hat{p}^K n^I (\widehat{(p^I)^{-1}})^2 \right]
\label{GammaOpr}
\end{eqnarray}
and the $\widehat{(p^I)^{-1}}$ are defined in equation (\ref{InvTriad}).
Note that $\hat{\Gamma}_I$ operators are diagonal in the triad basis
with eigenvalues denoted by $\Gamma_I(m_1, m_2, m_3)$.

In the presence of non-zero spin connection it is convenient to define a
new set of basis vectors:
\begin{eqnarray}
\widetilde{|m_1, m_2, m_3\rangle} 
& := & 
e^{ - \frac{1}{2}i \sum_I m_I \Gamma_I(m_1, m_2, m_3)}
|m_1, m_2, m_3\rangle
\end{eqnarray}

Denoting $|s\rangle = \tilde{s}_{m_1,m_2,m_3} \widetilde{|m_1,m_2,m_3\rangle}
\ , \ \hat{H} |s\rangle = 0$, leads to a partial difference equation.
This partial difference equation is quite complicated and is detailed in
\cite{Spin}. For the purposes of this paper we only need to look at the
{\em continuum approximation}. 

\section{Effective classical framework}
The effective classical Hamiltonian can be derived from the underlying
quantum equation in two steps. First one derives a {\em continuum
approximation} and in the second step derives an effective classical
Hamiltonian by considering a WKB ansatz for the continuum wave
function.

First, the fundamental difference equation is specialized for a slowly
varying discrete solution in the {\em continuum regime} defined by
$m_I := p^I/ (\frac{1}{2} \gamma\lP^2) \gg 1$. The leading terms are
then independent of the Barbero-Immirzi parameter $\gamma$ and lead to
the usual Wheeler--DeWitt equation for an interpolating continuum wave
function \cite{FundamentalDisc}.

Explicitly, for a slowly varying discrete wave function,
$\tilde{s}_{m_1, m_2, m_3} := \tilde{S}(p^I(m_I))$, $p^I(m_I) :=
\Case{1}{2}\gamma \lP^2 m_I$, the interpolating continuous wave
function $\tilde{S}$ satisfies the Wheeler--DeWitt equation
\cite{Spin},\footnote{In the equation (65) of this reference, the coefficient
of the first terms should be $1$ instead of $\Case{1}{4}$}
\[
\left[ 2 \ell_p^4 p^1 p^2 \frac{\partial^2 \sqrt{|p^1 p^2 p^3|} ~ 
\tilde{S}(p^1, p^2, p^3)}{\partial p^1 \partial p^2} ~ + ~ {\mbox{cyclic}}
\right]
 ~ + ~ W_{j,l}(p^1, p^2, p^3) \sqrt{|p^1 p^2 p^3|} ~ \tilde{S}(p^1, p^2, p^3) 
\]
\begin{equation} \label{WdWEq}
\hspace{3.0cm} = ~ - \kappa\,|p^1 p^2 p^3|^{\frac{3}{2}}
       \hat{\rho}^{\text{matter}}(p^1,p^2, p^3)\tilde{S}(p^1, p^2, p^3)
\,.
\end{equation}

Note that in the above equation, the $p^I$ dependence is obtained by
replacing $m_I/(2j) = \mu_I = p^I/(j \gamma \lP^2)$. This
will be important in the context of the potential and the matter terms
(see below).

In the second step, one takes the slowly varying wave function
$\tilde{T} := \sqrt{|p^1 p^2 p^3|} ~ \tilde{S}$ to be of the WKB form 
$\tilde{T} \sim e^{\text{i}A/\hbar}$. Then the real part of the
equation, to the leading order, $\hbar^0$, gives a partial differential
equation for the phase $A$, with the phase appearing only through its 
{\em first} derivatives with respect to $p^I$. Eq.~(\ref{CanonicalVar}) 
suggests the identification $p^I\Case{\partial A}{\partial p^I} := 
\Case{1}{2 \kappa} \pi_I$. Replacing the derivatives of the phase in
terms of $\pi_I$ in the partial differential equation converts it into 
an expression which has exactly the same form as the classical Hamiltonian 
(\ref{ClassHam}), except for the potential carrying the labels $j, l$. 
In short, we obtain a Hamilton-Jacobi equation for the phase. Classical
actions (corresponding to the Hamiltonian) along dynamical trajectories
are well known to solve the Hamilton-Jacobi equation. Thus constructing
slowly varying solutions of the Wheeler--DeWitt equation naturally leads
to analyzing the trajectories determined by the {\em classical
Hamiltonian} derived in the above manner.

The classicality of the Hamiltonian presumes that in (\ref{WdWEq}) 
the potential and the matter term are of order $\hbar^0$. 
{\em This feature however depends on the values of $p^I/(j \gamma
\lP^2)$}, because the ambiguity labels $j, l$ control the $p^I$
dependence of the potential. For instance, taking the value of $j$ to
be large, allows us to divide the $m_I \gg 1 $ continuum region into
(i) $j \gg m_I \gg 1$ and (ii) $m_I \gg j \gg 1$ sub-regions. Since
$F_l$ is a function of $m/(2j)$, the two sub-regions are more
conveniently demarcated as $p^I \ll j \gamma \lP^2$ and $p^I \gg j
\gamma \lP^2$ respectively.

In region (ii) the eigenvalues of the inverse triad operators correspond 
to the classical values (inverses of triad eigenvalues) and consequently the
spin connection acquires {\em no} dependence on $\gamma \lP^2$ when $\mu_I$ 
are expressed in terms of the $p^I$. The potential then reduces to the exact 
classical expressions without any dependence on $\gamma \lP^2$.

By contrast, in region (i) the $F_l(\mu_I)$ have a power law dependence
with {\em positive} power. This makes the spin connection have a
positive degree dependence on its arguments and introduces {\em
inverse} powers of $\gamma \lP^2$ in the potential (and matter) term
when $\mu_I$ are replaced by $p^I/(j \gamma \lP^2)$. These terms
clearly reflect {\em non-perturbative} quantum effects in region (i). 

Thus the argument that the effective Hamiltonian derived via the WKB
ansatz is {\em classical} ($\hbar^0$) is strictly valid only in region
(ii). Noting that in region (i), the inverse triads go to zero as 
$p^I \to 0 $, we will assume the validity of
the effective Hamiltonian in region (i) even though the potential has
inverse powers of $\gamma \lP^2$. In this manner, in an approximate
sense, we incorporate {\em non-perturbative} modifications of the
potential in an otherwise classical description.

The validity of the classical description is controlled by the
validity of the WKB approximation while how much of quantum
modifications can be seen in the classical framework is controlled by
$j$. Similar remarks apply to the matter Hamiltonian terms since the
matter density also typically gets quantum modifications from the inverse
volume operator.

The classical Hamiltonian deduced from the WKB approximation in the
above manner, including the extension to region (i), is what we will now 
refer to as {\em effective Hamiltonian}. Notice that imposition of the 
underlying quantum Hamiltonian constraint has automatically implied that 
the effective Hamiltonian must be zero which we interpret as the (modified) 
classical Hamiltonian constraint. 

To study the dynamical implications of the effective Hamiltonian constraint, 
recall first that in general relativity we can have two distinct
notions of dynamics or time evolution (see, e.g.,
\cite{KucharTime}). One is evolution with respect to an {\em external
time} which is a suitable coordinate in the space-time description of
the geometry and the other is evolution with respect to an {\em
internal time} which is one of the degrees of freedom treated as a
`clock'. The latter is obtained by solving the Hamiltonian constraint
for one suitably chosen momentum variable, say some $\pi_0$,
which becomes a function of the remaining phase space coordinates {\em and}
the conjugate coordinate $q^0$ identified as the internal time. The
evolution of the remaining degrees of freedom with respect to $q^0$ is
specified by the new Hamiltonian, ${\cal H} := - \pi_0$, which is in
general {\em time $(q^0)$ dependent}. Given a solution of the internal
dynamics, one can systematically construct a corresponding solution of
an external dynamics. What a quantum Hamiltonian constraint does is to
provide a {\em modified classical} Hamiltonian constraint -- our
effective Hamiltonian -- via the continuum approximation as outlined
above. This results in the modification of the trajectories of the
internal dynamics and correspondingly of the trajectories of the
external dynamics. In the context of homogeneous models, the internal
dynamics turns out to be time dependent but an {\em un-constrained}
system.

In the subsequent sections, we will analyze the internal dynamics determined 
from the modified Hamiltonian constraint.

\section{Internal time evolution}

We begin with the effective Hamiltonian (without matter) obtained from
the continuum approximation, 
\begin{equation} \label{Heff}
\kappa N H_{j,l}^{\rm eff} (\pi_I, q^I) ~ = ~ - \frac{1}{2} ( \pi_1 \pi_2
+ \pi_2 \pi_3 
+ \pi_3 \pi_1 ) + W_{j,l}(e^{2 q^1}, e^{2 q^2}, e^{2 q^3})
~~,~~\{q^I, \pi_J\} = \kappa \delta^I_J .
\end{equation}

Here, we are using dimensionless variables (or equivalently set the
unit of length such that $\gamma \lP^2 = 2$) and have put $m_I :=
e^{2 q^I} > 0$. The potential is given by
eq.(\ref{ModPotl}) with the spin connections given in
eq.(\ref{SpinCoeff}).

First we diagonalise the kinetic term in the Hamiltonian.
\begin{equation}
- \frac{1}{2} ( \pi_1 \pi_2 + \pi_2 \pi_3 + \pi_3 \pi_1 ) := -
  \frac{1}{4} \pi^T A \pi ~~ := -\frac{1}{4} \pi^T O D O^T \pi ~~ \text{where,}
\end{equation}

\begin{equation}
A := \left( \begin{array}{ccc}
0 & 1 & 1 \\
1 & 0 & 1 \\
1 & 1 & 0 \end{array} \right) ~~ = ~~ 
\left( \begin{array}{ccc}
\frac{2}{\sqrt{6}} & 0 & \frac{1}{\sqrt{3}} \\
\frac{-1}{\sqrt{6}} & \frac{1}{\sqrt{2}} & \frac{1}{\sqrt{3}} \\
\frac{-1}{\sqrt{6}} & -\frac{1}{\sqrt{2}} & \frac{1}{\sqrt{3}} 
\end{array} \right) 
\left( \begin{array}{ccc}
-1  & 0 & 0 \\
0 & -1  & 0 \\
0 & 0 & 2 \end{array} \right) 
\left( \begin{array}{ccc}
\frac{2}{\sqrt{6}} & -\frac{1}{\sqrt{6}} & -\frac{1}{\sqrt{6}} \\
0 & \frac{1}{\sqrt{2}} & -\frac{1}{\sqrt{2}} \\
\frac{1}{\sqrt{3}} & \frac{1}{\sqrt{3}} & \frac{1}{\sqrt{3}} \end{array} 
\right)
\end{equation}

\vskip 0.3cm

Defining $\pi' := O^T \pi, \ q' := O^T q$ we get the kinetic term as
$-\frac{1}{2}\pi_0^2 + \frac{1}{4}( \pi_+^2 + \pi_-^2 )$ and the
potential is obtained as a function of $q'=:(q^+, q^-, q^0)$ by
substituting $q = O q'$. Explicitly,
\begin{eqnarray}
q^1 ~ = ~  \frac{2}{\sqrt{6}} q^+ + \frac{1}{\sqrt{3}} q^0 
\mbox{\hspace{2.1cm}} 
& ~~~~~ & \mu_1 ~ = ~ \frac{1}{2j} e^{\frac{2}{\sqrt{3}}q^0} 
\ e^{\frac{4}{\sqrt{6}} q^+} 
\nonumber \\
q^2 ~ = ~  - \frac{1}{\sqrt{6}} q^+ + \frac{1}{\sqrt{2}} q^- + 
\frac{1}{\sqrt{3}} q^0 
& ~~~~~ & \mu_2 ~ = ~ \frac{1}{2j} e^{\frac{2}{\sqrt{3}}q^0} 
\ e^{-\frac{2}{\sqrt{6}} q^+} 
\ e^{\sqrt{2} q^-} 
\nonumber \\
q^3 ~ = ~  - \frac{1}{\sqrt{6}} q^+ - \frac{1}{\sqrt{2}} q^- + 
\frac{1}{\sqrt{3}} q^0 
& ~~~~~ & \mu_3 ~ = ~ \frac{1}{2j} e^{\frac{2}{\sqrt{3}}q^0} 
\ e^{-\frac{2}{\sqrt{6}} q^+} 
\ e^{- \sqrt{2} q^-} 
\end{eqnarray}

Clearly, $q^0 = \frac{1}{\sqrt{3}} (q^1 + q^2 + q^3) = \frac{1}{\sqrt{3}}
\ln (\text{volume})$. For an internal time description, this is a suitable
variable. The internal time description is obtained by solving 
$H^{\text{eff}}_{j,l} = 0$ for $\pi_0$:
\begin{equation}
- \pi_0 = {\cal{H}}(q^+, q^-, q^0, \pi_+, \pi_-) := \left[ \frac{1}{2} 
( \pi_+^2
  + \pi_-^2 ) + 2 W_{j,l}(q^+, q^-, q^0) \right]^{\frac{1}{2}} ~~ =:
  h^{\frac{1}{2}} 
\end{equation}

${\cal H}$ is to be used as the new Hamiltonian for analyzing the
behavior of $q^{\pm}$ as the volume shrinks to zero monotonically i.e.
as $q^0 \to - \infty$. Note that there are no longer constraints but the
new Hamiltonian is explicitly `time' dependent. We are also suppressing
the labels $j,l$ on ${\cal H}$ and $h$.

The equations of motion then take the form,
\begin{equation}
\frac{\md q^{\pm}}{\md q^0} = \frac{1}{2 \sqrt{h}} \pi_{\pm} ~~,~~
\frac{\md \pi_{\pm}}{\md q^0} = - \frac{1}{\sqrt{h}} 
\frac{\partial W_{j,l}}{\partial q^{\pm}}
\end{equation}
\begin{equation}
\frac{\md^2 q^{\pm}}{\md (q^0)^2} = - \frac{1}{2 h} \frac{\partial
W_{j,l}}{\partial q^{\pm}} - \frac{1}{h} \frac{\partial W_{j,l}}{\partial q^0} 
\frac{\md q^{\pm}}{\md q^0}
\end{equation}

Along a solution of the equation of motion, we can eliminate $h$ using
$h = 2 h ((\dot{q}^+)^2 + (\dot{q}^-)^2 ) + 2 W_{j,l}$ or $h = 2 W_{j,l}/(1 -
2((\dot{q}^+)^2 + (\dot{q}^-)^2))$. This leads to
\begin{equation} \label{Cartesian}
\frac{\md^2 q^{\pm}}{\md (q^0)^2} = - \left\{ 1 - 2\left(
\frac{\md q^+}{\md q^0}\right)^2 - 2 \left(\frac{\md q^-}{\md q^0}\right)^2
\right\} \left\{ \frac{1}{4 W_{j,l}} \frac{\partial
W_{j,l}}{\partial q^{\pm}} + \frac{1}{2 W_{j,l}} 
\frac{\partial W_{j,l}}{\partial q^0}
\frac{\md q^{\pm}}{\md q^0} \right\}
\end{equation}

It will also turn out to be convenient to introduce polar coordinates in the 
$q^{\pm}$ anisotropy plane. This is easily done:
Define $q^+ := R \cos(\theta)$, $q^- := R \sin(\theta)$. The 
conjugate momenta then are given by $\pi_+ := \pi_R \cos(\theta) - 
\pi_{\theta} \frac{\sin(\theta)}{R}$, $\pi_- := \pi_R \sin(\theta)
+ \pi_{\theta} \frac{\cos(\theta)}{R}$. The squared Hamiltonian,
$h$, and the equations of motion are given by
\begin{eqnarray}\label{HamR}
h(R, \theta, \pi_R, \pi_{\theta}, q^0) & = & \frac{1}{2}\left( \pi_R^2 + 
R^{-2} \pi_{\theta}^2 \right) + 2 W_{j,l}(p^I(q^0, R, \theta)) \ , \\
{\cal H}(R, \theta, \pi_R, \pi_{\theta}, q^0) & = & \sqrt{h(R, \theta, \pi_R,
\pi_{\theta}, q^0)} \; . 
\end{eqnarray}
Denoting the $\md/\md q^0$ by an over dot, the Hamilton's equations
of motion take the form,
\begin{eqnarray}\label{EOM}
\dot{\theta} = \frac{1}{2 \sqrt{h}} R^{-2} \pi_{\theta} & ,  & 
\dot{R} = \frac{1}{2 \sqrt{h}} \pi_R \ , \\
\dot{\pi_{\theta}} = -\frac{1}{\sqrt{h}} \left( 
\frac{\partial W_{j,l}}{\partial \theta} \right) & ,  &
\dot{\pi_R} = -\frac{1}{2 \sqrt{h}} \left( -R^{-3}
\pi_{\theta}^2 + 2 \frac{\partial W_{j,l}}{\partial R} \right) \ .
\end{eqnarray}
The second order equations are obtained as,
\begin{eqnarray}
\ddot{R} - R \dot{\theta}^2 & = & -
h^{-1}\left( \frac{1}{2} \frac{\partial W_{j,l}}{\partial R} + 
\dot{R} \frac{\partial W_{j,l}}{\partial q^0} \right) \ ,
\label{Radial}\\
\ddot{\theta} + \frac{2}{R} \dot{R} \dot{\theta}
 & = & -
h^{-1}\left( \frac{1}{2} R^{-2} \frac{\partial W_{j,l}}{\partial \theta} + 
\dot{\theta} \frac{\partial W_{j,l}}{\partial q^0} \right)
\ . \label{Angular} 
\end{eqnarray}

Along the dynamical trajectories, we can eliminate the conjugate momenta
from the Hamiltonian and obtain,
\begin{equation}\label{HamCond}
h^{-1} = \frac{ 1 - 2 \left( \dot{R}^2 + R^2 
\dot{\theta}^2 \right) }{2 W_{j,l}} ~~ > 0 ~~~ (\text{strict inequality}) \ .
\end{equation}

When the potential is zero, as in Bianchi I, the right hand sides of
(\ref{Radial}, \ref{Angular}) are zero. It is easy to see that
the Kasner solution is obtained as $\theta = \text{constant}$, $R(q^0) = -
\frac{1}{\sqrt{2}} q^0 + \text{constant}$. When the potential is
non-zero, the equations are quite complicated. To proceed with the
qualitative analysis, we will study the effective potential in the next
section.

\section{Study of the Effective Potential}
\label{WithinDomain}

The effective potential is more conveniently studied in terms of the
polar variables. The following expressions are useful for our subsequent
analysis.
\begin{eqnarray}
q^+ & := & R \ \cos \theta \hspace{3.6cm}
q^- ~ := ~ R \ \sin \theta \nonumber \\
Q(j, q^0) & := & \frac{e^{\frac{2}{\sqrt{3}} q^0}}{2 j} \hspace{0.0cm} = ~~
\frac{(\text{volume})^{2/3}}{2j} \\
\mu_1 & = & Q e^{\frac{4}{\sqrt{6}} R \cos \theta } 
~~~~~~~~~~~~~~~~ \hspace{1.6cm} = ~~~ Q e^{+ \frac{4}{\sqrt{6}} R \cos \theta }
\nonumber \\
\mu_2 & = & Q e^{-\frac{2}{\sqrt{6}} R \cos \theta (1 -
\sqrt{3} \tan \theta )} ~~~ \hspace{1.5cm} = ~~~ Q e^{- \frac{4}{\sqrt{6}} R
\cos(\theta + \frac{\pi}{3}) } \label{MuDefs} \\
\mu_3 & = & Q e^{-\frac{2}{\sqrt{6}} R \cos \theta (1 +
\sqrt{3} \tan \theta ) } ~~~ \hspace{1.5cm} = ~~~ Q e^{- \frac{4}{\sqrt{6}} R
\cos(\theta - \frac{\pi}{3}) } \nonumber 
\end{eqnarray}
Clearly, $\mu_1 \mu_2 \mu_3  =  Q^3$ is independent of the $q^{\pm}$.

Using the obvious abbreviation, $F_I := F_l(\mu_I)$, (label $l$ on $F$
is suppressed) the potential is given explicitly by,
\begin{eqnarray} \label{PotDetail}
\frac{W_{j,l}(\vec{\mu})}{ 2 j^2} & = & \left[ -2 \left\{
\mu_1 \mu_2(\mu_1 F_2 + \mu_2 F_1) + 
\mu_2 \mu_3(\mu_2 F_3 + \mu_3 F_2) + 
\mu_3 \mu_1(\mu_3 F_1 + \mu_1 F_3) 
\right\} + \right. \nonumber \\
& & 3 \left\{
(\mu_1 \mu_2 F_3)^2 + (\mu_2 \mu_3 F_1)^2 + (\mu_3 \mu_1 F_2)^2 \right\}
+ 3 Q^3 \left\{ \mu_1 F_2 F_3 + \mu_2 F_3 F_1 + \mu_3 F_1 F_2 \right\}
\nonumber \\ 
& & - Q^3 \left\{
\mu_1 \mu_2 F_3 ( F_1^2 + F_2^2 ) + 
\mu_2 \mu_3 F_1 ( F_2^2 + F_3^2 ) + 
\mu_3 \mu_1 F_2 ( F_3^2 + F_1^2 ) \right\} \nonumber \\
& & \left. - 2 Q^3 \left\{
\mu_1 \mu_2 F_3^3 + \mu_2 \mu_3 F_1^3 + \mu_3 \mu_1 F_2^3 \right\}
+ Q^6 \left\{ F_1^2 F_2^2 + F_2^2 F_3^2 + F_3^2 F_1^2 \right\} \right]
\end{eqnarray}

We define the {\em far regime}, by $\mu_I \gg 1$ and the {\em near
regime} by $0 < \mu_I \ll 1$. Note that the regimes are delineated in terms
of the $\mu_I$ and {\em not} in terms of the anisotropy variables
$q^{\pm}$. The regimes are realized for any bounded region in
the anisotropy plane ($q^+$-- $q^-$ plane) and sufficiently {\em large
volume} ($Q \gg 1$) and sufficiently {\em small volume} ($Q \ll 1$)
respectively. In these regimes, the potential simplifies to the forms,
\begin{eqnarray}
W_{j,l}^{\text{near}}(\vec{\mu}) & \approx & 
- 4 j^2 \left(\frac{3}{l+1}\right)^{\frac{1}{1-l}} \left[~
\mu_3^{\frac{2-l}{1-l}} ~ (\mu_1^2 + \mu_2^2) ~ + ~ 
\mu_1^{\frac{2-l}{1-l}} ~ (\mu_2^2 + \mu_3^2) ~ + ~ 
\mu_2^{\frac{2-l}{1-l}} ~ (\mu_3^2 + \mu_1^2) ~\right] \nonumber \\
& = & -2 (2j)^{-\frac{2-l}{1-l}} \left(\frac{3}{l+1}\right)^{\frac{1}{1-l}}
e^{\frac{2}{\sqrt{3}} q^0(\frac{4-3l}{1-l})} \left[ \ 
e^{-\frac{2}{\sqrt{6}}q^+(\frac{4-3l}{1-l})}\cosh\left(\frac{l}{1-l} 
\sqrt{2} q^-\right) \right. \nonumber \\
& & \left. \mbox{\hspace{6.0cm}} ~ + ~ 
e^{\frac{2}{\sqrt{6}}q^+(\frac{2-3l}{1-l})} \cosh\left( \frac{2-l}{1-l} 
\sqrt{2} q^- \right)
\right. \nonumber \\
& & \left. \mbox{\hspace{6.0cm}} +~ e^{\frac{4}{\sqrt{6}}q^+(\frac{1}{1-l})} 
\cosh\left(2 \sqrt{2} q^-\right) \ \right] \label{PotNear} \\
& & \nonumber \\
W_{j,l}^{\text{far}}(\vec{\mu}) & \approx & 
2 j^2 \left[
\left(\frac{\mu_1 \mu_2}{\mu_3}\right)^2 +
\left(\frac{\mu_2 \mu_3}{\mu_1}\right)^2 +
\left(\frac{\mu_3 \mu_1}{\mu_2}\right)^2 -
2 \left( \ \mu_1^2 + \mu_2^2 + \mu_3^2 \ \right) \ \right] \nonumber \\
& \approx & \frac{1}{2} e^{\frac{4}{\sqrt{3}} q^0} \left[\ 
e^{-\frac{16}{\sqrt{6}}q^+} ~-~ 4 e^{- \frac{4}{\sqrt{6}}q^+}
\cosh\left( 2 \sqrt{2} q^- \right) \right. \nonumber \\
& & \left. \hspace{3.0cm} 
+~ 2 e^{\frac{8}{\sqrt{6}}q^+}\left\{\cosh\left(4 \sqrt{2}
q^-\right) - 1 \right\} \ \right] \label{PotFar}
\end{eqnarray}

Comparison of expression (\ref{PotFar}) with Misner's expressions 
\cite{Mixmaster} suggests the identification: $q^+ \to \frac{\sqrt{6}}{2} 
\beta_+ \ , q^- \to \frac{\sqrt{6}}{2} \beta_- \ , q^0 \to -\sqrt{3}
\Omega$. We will continue to work with the $q$'s.

Note that although in the two limiting cases the $q^0$ dependence
factors out in the potential, this is {\em not} true in general. This is
different from the classical potential which is just the far form of the
effective potential. 

The factorization property of the far and near potential allows to
show that there can be no bounce in the corresponding regimes, i.e.,
once the volume decreases, it must keep on decreasing until it reaches
zero \cite{BianchiIXOsc}. This follows from the external time dynamics
generated by the constraint
$H=-\pi_0^2+\frac{1}{2}(\pi_+^2+\pi_-^2)+2W=0$. For $q^0$ and its
conjugate the Hamiltonian equations of motion are $\dot{q}^0=-2\pi_0$
and $\dot{\pi}_0=-\partial H/\partial q^0= -2\partial W/\partial
q^0=-2kW$ with positive constants $k$ in the near and far regime. This
implies
\[
 \frac{\md}{\md
 t}\left(\pi_0e^{-kq^0/2}\right)=ke^{-kq^0/2}(\pi_0^2-2W)=
 ke^{-kq^0/2}(\pi_+^2+\pi_-^2)/2\geq 0\,.
\]
Thus, once the volume decreases, i.e.\ $\dot{q}^0<0$ and $\pi_0>0$, it
cannot turn around since $\pi_0e^{-kq^0/2}$ starts out positive
and cannot decrease. This proof, however, relies on the factorization
of the potential into a part depending exponentially on $q^0$ and a
part depending only on the anisotropies. In regions where such a
factorization is not possible, it is not clear if bounces or
oscillations are ruled out. The modification of the potential could
lead to a bounce at positive volume as observed in \cite{BounceClosed}
for a closed isotropic model with scalar matter. If a similar bounce
would occur in the effective Bianchi IX dynamics, the system would
clearly not show the classical chaotic behavior. Thus, we can assume
here that there is no bounce in order to investigate the potentially
chaotic behavior.

Other asymptotic regions correspond to two of the $\mu_I$'s large
(small) and the third one small (large). These are realized for
example, for a fixed and moderate $Q$ and large $R$. The behavior of
$\mu_I$'s is then controlled by the angles. These regions divide the
anisotropy plane into {\em angular sectors} with $R$ large. In these
regions, we could use the asymptotic forms of $F_l(\mu_I)$'s and
obtain the corresponding approximate forms of the potential. These
forms are of course {\em different} from the near and far forms
displayed in equations (\ref{PotNear}, \ref{PotFar}). These are
discussed in more details in section \ref{Extrapolation}.

\subsection{Approximate wall picture}

One can see the qualitative behavior of the potential (classical or
effective) by specializing to $q^- = 0$ or $\theta = 0\, \text{and} \, 
\pi$. For this $q^-$, we get $\mu_2 = \mu_3 := \mu$ and $\mu_1 := \nu = 
\frac{Q^3}{\mu^2}$. It follows that $\Gamma_2 = \Gamma_3$ and the potential 
simplifies a bit. A wall, i.e.\ positive potential, is seen for $\mu \gg 1$ 
so that $F_l(\mu) \approx \mu^{-1}$ may be used but $\nu$ is not so small 
that $F_l(\nu) \approx 4 \nu^2$ can be used. Making this substitution for 
$F_l(\mu)$ one obtains,
\begin{eqnarray}
\Gamma_3 ~ = ~ \Gamma_2 & \approx & \frac{1}{2} \mu F_l(\nu) ~~~~,~~~~ 
\Gamma_1 ~ \approx ~ 1 - \frac{1}{2} \mu^2 F_l^2(\nu) ~~ ;\nonumber \\
W_{j,l}(\nu, \mu, \mu) & \approx & 2 j^2 \mu^4 F_l^2(\nu) \left\{ 3 - 2 \nu
F_l(\nu) \right\} - 8 j^2 \mu^2 \nonumber \\
& \approx & 2 j^2 \frac{Q^6}{\nu^2} F_l^2(\nu) \left\{ 3 - 2 \nu
F_l(\nu) \right\} - 8 j^2 \frac{Q^3}{\nu} \ , \label{ExactWall} \\
& =: & V_l(Q, \nu) - 8 j^2 \frac{Q^3}{\nu}
\hspace{2.0cm}\text{(definition of $V_l(Q, \nu)$)}\ .\label{Wall}
\end{eqnarray}
We refer to $W_{j,l}(\nu, \mu, \mu)$ in eq.(\ref{ExactWall}) as the {\em
wall potential} and to $V_l(Q, \nu)$ defined in (\ref{Wall}) as the
{\em approximate wall potential}.

The last term in the wall potential can be neglected for $\nu \sim o(1)$
and $Q^3/\nu = \mu^2 \gg 1$, since $F^2_l(\nu)( 3 - 2 \nu F_l(\nu) )$ is 
of the order of $1$ for this range of values. Neglecting this term and 
substituting $Q^6 = (\text{volume}) ^4/(2 j)^6$ in the approximate wall
potential, gives the expression quoted in \cite{NonChaos}. 

For $\nu \gg 1$, one can replace $F_l(\nu)$ by $\nu^{-1}$ as well leading 
to $V_l(Q, \nu) \approx 2 j^2 Q^6/\nu^4 \approx
\frac{1}{2}\text{(volume)}^{4/3} e^{- \frac{16}{\sqrt{6}} q^+} $ matching 
with the {\em leading} term of eq.(\ref{PotFar}) for $q^- = 0$ which is 
the {\em classical wall potential}.

In contrast to the classical wall, the wall of the effective potential
has finite height. The location of its maximum along the $q^+$ axis is
defined by $\frac{\partial V_l}{\partial q^+} = 0$. Its solution will
define a function $q^+(Q)$. The motion of this position is obtained by
solving $\frac{\md}{\md Q}(\frac{\partial V_l}{\partial \nu}) = 0$. Recall
$\nu = Q e^{\frac{4}{\sqrt{6}} q^+} =: Q e^x$ such that
\begin{eqnarray}
\frac{\partial V_l(Q, \nu(Q,x))}{\partial x} & = & 0 ~~~~
\Longleftrightarrow ~~~~ \frac{\partial V_l}{\partial \nu} ~ = ~ 0
\nonumber \\
\frac{\partial^2 V_l}{\partial x^2} & < & 0 ~~~~
\Longleftrightarrow ~~~~ \frac{\partial^2 V_l}{\partial \nu^2} ~ < ~ 0
\end{eqnarray}
Thus, a (local) maximum of the approximate wall potential can be
defined either in terms of derivatives with respect to $x =
\frac{4}{\sqrt{6}} q^+$ or with respect to $\nu$.

Notice that for the classical wall, there is no local extremum and
therefore one has to track the motion of a fixed value of the potential.
Thus, when the volume decreases, $q^+$ must also {\em decrease},
i.e.\ the wall moves away from the isotropy neighborhood (note that the
wall is seen for $\mu\gg 1$ which implies $q^+<0$). By contrast, the
approximate wall potential does have a local maximum at some $\nu_0$.
Fixed $\nu$ implies that, with decreasing volume, the {\em height} of
the approximate wall decreases {\em and} $q^+$ {\em increases}. Thus
the approximate wall moves towards the isotropy neighborhood with
decreasing height. Eventually of course the wall potential must be
used and the height actually crosses zero for a certain volume depending
upon the values of $j, l$.

It is difficult to obtain a convenient approximation for a wall along an
arbitrary direction and numerics must be used. Numerically, the full 
effective  potential becomes zero everywhere, in the anisotropy plane, 
for $Q \approx 1.08511$ or for the dimensionless volume of about 
$(2.172 j)^{3/2}$ (for the choice $l = \Case{1}{2}$).

From these studies, it is apparent that the principal reason for the usual 
chaotic behavior disappears suggesting that the usual `Kasner epochs' will 
continue without any replacements once the volume becomes comparable to the 
Planck volume. This in turn suggests that the Kasner solution should be 
stable under the effective potential.

\subsection{Stability of Kasner solution}

This is most easily done in the Cartesian form
(\ref{Cartesian}). Substitute $q^{\pm}(q^0) = K^{\pm}(q^0) + \delta
q^{\pm}(q^0)$ where $K^{\pm} := - \frac{1}{\sqrt{2}} u^{\pm} q^0, ~
u^+ := \cos(\phi), u^- := \sin(\phi)$. Linearization gives,
\begin{eqnarray}
\ddot{\delta q}^{\pm} & = & - \left\{ -4 \dot{K}^+ \dot{\delta q}^+ -4
\dot{K}^- \dot{\delta q}^- \right\} \left. \left\{ \frac{1}{4 W_{j,l}} 
\frac{\partial
W_{j,l}}{\partial q^{\pm}} + \dot{q}^{\pm} \frac{1}{2 W_{j,l}} 
\frac{\partial
W_{j,l}}{\partial q^0} \right\}\right|_{\text{Kasner}} \nonumber \\
\label{Stability}
 & = & - 2 \sqrt{2} \left\{ \cos\phi \dot{\delta q}^+ +
\sin\phi \dot{\delta q}^- \right\} \left. \left\{ \frac{1}{4
W_{j,l}} \frac{\partial
W_{j,l}}{\partial q^{\pm}} - \frac{u^{\pm}}{2 \sqrt{2}} \frac{1}{W_{j,l}} 
\frac{\partial
W_{j,l}}{\partial q^0} \right\}\right|_{\text{Kasner}} \nonumber \\
 & := & \kappa^{\pm}(q^0, \phi) \left\{ \cos\phi \dot{\delta q}^+ +
\sin\phi \dot{\delta q}^- \right\} ~~~~~~ \Rightarrow\\
\frac{\md}{\md q^0} \left( 
\begin{array}{c} 
\dot{\delta q}^+ \\
\dot{\delta q}^- 
\end{array}
\right) & = & 
\left( \begin{array}{cc}
\kappa^+ \cos\phi & \;\;\kappa^+ \sin\phi \\
\kappa^- \cos\phi & \;\;\kappa^- \sin\phi 
\end{array} \right) 
\left( \begin{array}{ c } 
\dot{\delta q}^+ \\
\dot{\delta q}^- 
\end{array} \right)
\end{eqnarray}
It is obvious now that the determinant of the matrix is zero and the only
(possibly) non-zero eigenvalue is given by the trace: $\lambda := \kappa^+ 
\cos\phi + \kappa^- \sin\phi$. 

The eigenvector corresponding to the zero eigenvalue is given by
$\dot{\delta q^-} = - \cot\phi \dot{\delta q^+}$. The first braces
in the middle equation of (\ref{Stability}) then vanishes. This zero mode
implies that the perturbed solution is also a Kasner solution. But we
know that the dynamical equations do {\em not} admit Kasner solution
since the potential is generally non-zero and $h$ is to be {\em finite
and positive} and this solution of the perturbed equation must be
rejected.

Thus the Kasner solution is stable if $\lambda > 0$ (positive because 
we are considering the backward evolution, decreasing $q^0$ ). Up to here, 
no specific form of potential has been used.

To compute $\kappa^{\pm}$ we need to compute the potential on the Kasner
solution. Substituting Kasner $K^{\pm}(q^0)$ in the $\mu_I$ gives,
\begin{eqnarray}\label{MuKasner}
2 j \mu_1 & = & e^{\frac{2}{\sqrt{3}} q^0 ( 1 - \cos\phi )}
\nonumber \\
2 j \mu_2 & = & e^{\frac{2}{\sqrt{3}} q^0 ( 1 + \frac{1}{2}\cos\phi -
\frac{\sqrt{3}}{2}\sin\phi )} \\
2 j \mu_3 & = & e^{\frac{2}{\sqrt{3}} q^0 ( 1 + \frac{1}{2}\cos\phi +
\frac{\sqrt{3}}{2}\sin\phi )} \nonumber
\end{eqnarray}
It is obvious now that all the exponents are non-negative over all $\phi$ 
implying that all $\mu_I$ will (generically) {\em decrease} as $q^0$ is 
decreased. Only for $\phi = 120^0$ where $\mu_2 = 1/(2j)$, $\phi = 240^0$ 
when $\mu_3 = 1/(2j)$ and $\phi = 0^0$ where $\mu_1 = 1/(2j)$, all
exponents are positive. By taking $j > 1/2$ we can ensure that all the
$\mu_I$ are much less than 1 at some initial $q^0$. We can then use the 
near-form of the potential to compute the derivatives.

Defining, generally, $2 j \mu_I =: e^{\nu_I(q^0, q^{\pm})}$ one can
obtain the derivatives of the potential as,
\begin{eqnarray}
\delta W_{j,l} & = & \sum_I \delta \nu_I \mu_I 
\frac{\partial W_{j,l}}{\partial
\mu_I} ~~~~~~~~~~~ \Rightarrow \\
\frac{\partial W_{j,l}}{\partial q^0} & = & \frac{2}{\sqrt{3}} \sum_I
\mu_I \frac{\partial W_{j,l}}{\partial \mu_I} \nonumber \\
\frac{\partial W_{j,l}}{\partial q^+} & = & \frac{2}{\sqrt{6}} \left(
2 \mu_1 \frac{\partial }{\partial \mu_1} 
- \mu_2 \frac{\partial }{\partial \mu_2} 
- \mu_3 \frac{\partial }{\partial \mu_3} \right) W_{j,l} \\
\frac{\partial W_{j,l}}{\partial q^-} & = & \sqrt{2} \left(
\mu_2 \frac{\partial }{\partial \mu_2} 
- \mu_3 \frac{\partial }{\partial \mu_3} \right) W_{j,l} \nonumber \\
W_{j,l}^{\text{near}} & = & -4 j^2 
 \left(\frac{3}{l+1}\right)^{\frac{1}{1-l}}\left[
\mu_1^L (\mu_2^2 + \mu_3^2) + 
\mu_2^L (\mu_3^2 + \mu_1^2) + 
\mu_3^L (\mu_1^2 + \mu_2^2) \right]
\end{eqnarray}
where $L=\frac{2-l}{1-l}$.
The logarithmic derivatives of the near potential are given by, 
\begin{eqnarray}
\frac{1}{W^{\text{near}}_{j,l}}
 \frac{\partial W^{\text{near}}_{j,l}}{\partial q^0} 
& = & \frac{2 (L + 2)}{\sqrt{3}} \hspace{3.0cm} L ~ := \frac{2 - l}{1 - l}
 \label{ZeroDiff}\\
\frac{1}{W^{\text{near}}_{j,l}}
\frac{\partial W^{\text{near}}_{j,l}}{\partial q^+} 
& = & \frac{2}{\sqrt{6}} 
\left[
\mu_1^L (\mu_2^2 + \mu_3^2) + 
\mu_2^L (\mu_3^2 + \mu_1^2) + 
\mu_3^L (\mu_1^2 + \mu_2^2) \right]^{-1} 
\left[
2 (L - 1) \mu_1^L(\mu_2^2 + \mu_3^2) \right. \nonumber \\
& & \left. +
\mu_2^L(- (L + 2) \mu_3^2 + (4 - L) \mu_1^2) +
\mu_3^L((4 - L) \mu_1^2 - (L + 2) \mu_2^2) \right]
\label{PlusDiff}\\
\frac{1}{W^{\text{near}}_{j,l}}
\frac{\partial W^{\text{near}}_{j,l}}{\partial q^-} 
& = & \sqrt{2}
\left[
\mu_1^L (\mu_2^2 + \mu_3^2) + 
\mu_2^L (\mu_3^2 + \mu_1^2) + 
\mu_3^L (\mu_1^2 + \mu_2^2) \right]^{-1}  
\left[
2 \mu_1^L(\mu_2^2 - \mu_3^2) \right. \nonumber \\
& & \left. 
+ \mu_2^L((L - 2) \mu_3^2 + L \mu_1^2) -
\mu_3^L( L\mu_1^2 + (L - 2)\mu_2^2) \right] \label{MinusDiff}
\end{eqnarray}
These now have to be evaluated on the Kasner solution. 

Observe that the logarithmic derivatives of the potential with respect
to $q^{\pm}$ are ratios of homogeneous polynomials of the same degree.
Hence any common factors in $\mu_I$ will cancel out. From
(\ref{MuKasner}), it is obvious that effectively we can take,
\begin{equation}\label{EffectiveMu}
\mu_1 := \mu^{-3}, ~ \mu_2 := \nu^{-1}, ~ \mu_3 := \nu,
~~~{\text{where}}~~ \mu := e^{\frac{q^0}{\sqrt{3}} \cos\phi }, 
~ \nu := e^{q^0 \sin\phi} \, .
\end{equation}
The common factor canceled is $Q(q^0, j) \mu$. Substituting these in
(\ref{PlusDiff}, \ref{MinusDiff}) leads to,

\begin{eqnarray}
\frac{1}{W_{j,l}^{\text{near}}}\frac{\partial W_{j,l}^{\text{near}}}
{\partial q^+} & = & \frac{2}{\sqrt{6}} 
\left[
\mu^{-3L} (\nu^2 + \nu^{-2}) + 
\mu^{-6} (\nu^L + \nu^{-L}) + 
(\nu^{L-2} + \nu^{-(L-2)}) 
\right]^{-1} \nonumber \\ 
& & \left[
2(L-1) \mu^{-3L} (\nu^2 + \nu^{-2}) + 
(4-L)\mu^{-6} (\nu^L + \nu^{-L}) - \right. \nonumber \\ 
& & \hspace{4.9cm} \left. (L+2) (\nu^{L-2} + \nu^{-(L-2)}) 
\right]
\nonumber \\
& =: & \frac{2}{\sqrt{6}} D^{-1} a_+ \label{PlusDiffSimp}\\
\frac{1}{W_{j,l}^{\text{near}}}\frac{\partial W_{j,l}^{\text{near}}}
{\partial q^-} & = & - \sqrt{2}
\left[
\mu^{-3L} (\nu^2 + \nu^{-2}) + 
\mu^{-6} (\nu^L + \nu^{-L}) + 
(\nu^{L-2} + \nu^{-(L-2)}) 
\right]^{-1} \nonumber \\
& & \left[
2 \mu^{-3L} (\nu^2 - \nu^{-2}) + 
L \mu^{-6} (\nu^L - \nu^{-L}) + 
(L-2)(\nu^{L-2} - \nu^{-(L-2)}) 
\right] \nonumber \\
& =: & - \frac{2}{\sqrt{6}} D^{-1} a_- \label{MinusDiffSimp} \\
\kappa^+ & = & \frac{2 (L + 2)}{\sqrt{3}} \cos\phi -
\frac{1}{\sqrt{3}} \frac{a_+}{D}  \\
\kappa^- & = & \frac{2 (L + 2)}{\sqrt{3}} \sin\phi +
\frac{1}{\sqrt{3}} \frac{a_-}{D}  \\
\lambda & = & \frac{1}{\sqrt{3}}\left[ 2 (L + 2) - \frac{a_+\cos\phi}{D}
 + \frac{a_-\sin\phi}{D} \right\} \label{Lambda} 
\end{eqnarray}
The eigenvalue $\lambda$ is a function of $\phi$ and $q^0$. Generic 
stability of Kasner solution requires $\lambda$ to be positive for {\em all
$\phi$} and at {\em every $q^0$ less than the initial $q^0$}. A necessary 
condition for this to be true is that  $a_+^2 + a_-^2 < 4 (L + 2)^2 D^2$. 
This is easily checked independently of $\phi, q^0$, completing the proof
that the Kasner solution is stable under inclusion of the effective potential
for all $\phi$ and for all $ q^0$ less than the initially chosen $q^0$.

An identical calculation can be repeated for the classical potential
by using the far form of the potential (which is independent of $l$,
of course). However, now $\lambda$ {\em does not have a definite sign
for all $\phi, q^0$.}  For a fixed $\phi$, $\lambda$ changes sign. The
positive sign corresponds to `Kasner epochs' while the negative sign
corresponds to `replacement of Kasner epochs'.

\subsection{Relation to Kasner Sphere}

It is conventional to analyze the asymptotic behavior of the approach to
a singularity in terms of the scale factors of the classical geometry. We
can obtain such a description by noting that (all scale factors, triad
components positive),
\begin{eqnarray}
p^I & = & a_J a_K \\
a_I & = & \frac{\text{volume}}{p^I} ~~~~ 
~ = ~ \left(2j\right)^{\frac{1}{2}} \frac{
Q^{3/2}} {\mu_I} ~~~~~~ \text{i.e.} \\
a_1 & = & e^{\frac{q^0}{\sqrt{3}}}
\ e^{- \frac{4}{\sqrt{6}}\ R \cos \, \theta } \\
a_2 & = & e^{\frac{q^0}{\sqrt{3}}}
\ e^{+ \frac{2}{\sqrt{6}}\ R \cos \, \theta ( 1 - \sqrt{3}
\tan\, \theta ) } \\
a_3 & = & e^{\frac{q^0}{\sqrt{3}}}
\ e^{+ \frac{2}{\sqrt{6}}\ R \cos \, \theta ( 1 + \sqrt{3}
\tan\, \theta ) } 
\end{eqnarray}
For the Kasner solution, $R(q^0) = - \frac{1}{\sqrt{2}}q^0 , \theta =
\theta_0$ while the standard parameterization is $a_I \sim \text{t}^{2
\alpha_I} = e^{2 \alpha_I \ln \text{t}}$ such that $\sum_I \alpha_I =
1 = \sum_I \alpha_I^2$ (`Kasner sphere').  Putting $q^0 = 2 \zeta \ln
\text{t}$, comparison implies,
\begin{eqnarray}
\alpha_1 & = &  \frac{\zeta}{\sqrt{3}} \left\{1 + 2 \cos\,
\theta_0\right\} \nonumber \\
\alpha_2 & = &  \frac{\zeta}{\sqrt{3}} \left\{1 - \cos\,
\theta_0 
(1 - \sqrt{3} \tan\, \theta_0) \right\} \nonumber \\
\alpha_3 & = &  \frac{\zeta}{\sqrt{3}} \left\{1 - \cos\,
\theta_0 
(1 + \sqrt{3} \tan\, \theta_0) \right\}
\end{eqnarray}

The Kasner Sphere conditions are then automatically satisfied with the
choice $\zeta = \frac{1}{\sqrt{3}}$ which identifies the co-moving time
t as $e^{\frac{\sqrt{3}}{2} q^0}$ which is consistent with the
definition of $q^0$.

It is obvious now that a point on the Kasner sphere corresponds to a
choice of $\theta_0$. Oscillatory behavior corresponds to a motion of
the system point passing through different angular sectors mentioned above.

Observe that a deviation from Kasner solution ($\theta$ non-constant)
does {\em not} imply oscillatory behavior. The $\theta$ motion may
still be confined to one of the angular sectors leading to a {\em
Kasner-like} behavior. For the classical potential with its confining
nature (infinite walls), non-constant $\theta$ implies traversing all
the angular sectors and thus oscillatory behavior. For the modified
potential such an implication is {\em not} automatic although conceivable.

In the next section on the qualitative analysis we will see that the
asymptotic solution is indeed a Kasner solution. 

\subsection{Domain of Validity}

Before we consider the qualitative analysis, it is useful to make
explicit the domain of validity of the effective Hamiltonian. Recall that 
the derivation of the effective Hamiltonian is based on the validity of
the continuum approximation and the validity of the WKB approximation.

The continuum approximation itself is valid for $m_I \gg 1$ which implies
$\mu_I \gg \epsilon := 1/(2j)$. In terms of the $q^{\pm, 0}$ these
restrictions translate to:
\begin{eqnarray}
Q \gg \epsilon ~~ & \Leftrightarrow & q^0 > 0 \ ; 
\nonumber \\
\frac{\sqrt{6}}{4} \ln (\epsilon/Q) & \ll q^+ \ll &
        \frac{\sqrt{6}}{2} \ln (Q/\epsilon) \ ; \\
- \frac{1}{\sqrt{3}}\left( \frac{\sqrt{6}}{2} \ln (Q/\epsilon) - q^+ \right) 
& \ll  q^-  \ll &
\frac{1}{\sqrt{3}}\left( \frac{\sqrt{6}}{2} 
\ln (Q/\epsilon) - q^+ \right) \ .\nonumber
\end{eqnarray}

Equivalently, using $Q/\epsilon = \text{Vol}^{2/3}$ we have,
\begin{eqnarray}
\text{Vol} \gg 1 && \ ; \nonumber \\
- \frac{1}{\sqrt{6}} \ln (\text{Vol}) & \ll q^+ \ll &
\sqrt{\frac{2}{3}} \ln (\text{Vol}) \ ; \\
- \frac{1}{\sqrt{3}} \left(\sqrt{\frac{2}{3}} \ln (\text{Vol}) - q^+ \right)
& \ll q^- \ll & 
\frac{1}{\sqrt{3}} \left(\sqrt{\frac{2}{3}} \ln (\text{Vol}) - q^+ \right) \ .
\end{eqnarray}

For a given $l$, the effective potential becomes negative for some
$Q^*_l$, independently of $j$ and the corresponding $ \text{vol}^*_l
\approx (2j Q^*_l)^ {\frac{3}{2}}$. To be consistent with the domain
of validity, $\text{Vol}^*_l > 1$ must hold which can be arranged by
choosing $j$ sufficiently large.  Having chosen $\text{Vol}^*_l$, the
limits on $q^{\pm}$ follow.

Some typical values are: $Q^*_{0.1} \approx 0.91 \ ; Q^*_{0.5} \approx 1.086
\ ; Q^*_{0.9} \approx 1.38$. Clearly, for smaller $l$ we need to choose
a larger $j$ in order to stay within the domain of validity. (In this
context it is interesting to note that both conceptual considerations
in the full theory and phenomenological studies in the context of
early universe inflation point to values $j>1/2$ \cite{Robust}.) Thus
we see that disappearance of the walls can be seen generically and
reliably within the effective Hamiltonian picture. For the plots shown
in \cite{NonChaos}, the ranges of $q^{\pm}$ are in the domain of
validity.

For the stability analysis above, for large enough $j$, we can have all
$\mu_I \ll 1$ even though $q^0 > 0$ justifying the use of the near form
of the potential. The Kasner stability is thus seen staying consistent with
the validity of the effective Hamiltonian. 

One reason for this demarcation is that for any given fixed volume, the 
large $R$ form of the effective potential is very different from that of the
classical potential. In particular, the classical potential is always large
and positive while the effective one is large and negative. Only the inner
side of the walls of the effective potential `matches' with the classical
potential for large volume. Consequently, the `initial conditions' for the
internal dynamics can be either in the inner side of the walls or the outer
side of the walls -- an option not available for the classical potential.

In the next section, we will go beyond the domain of validity by considering 
the limit $\text{Vol} \to 0$.

\section{Qualitative analysis}
\label{Extrapolation}

In this section we will pretend that the effective description is
valid all the way to zero volume or $q^0 \to - \infty$. Although we
know that this cannot be true, we are considering this by way of
contrasting the dynamics implied by the classical potential. We will
therefore be interested in the asymptotic behavior as $q^0 \to -
\infty$ of the trajectories in the anisotropy plane. This analysis
proceeds in stages. Asymptotically one of the following two
possibilities can happen: either $R(q^0)$ remains bounded or it does
not. In the first stage, we will first rule out a bounded asymptotic
motion while in the second stage we will show that asymptotic motion
always approaches a Kasner trajectory.

\begin{prop} \label{PropBdd}
Asymptotically, $R(q^0)$ cannot remain bounded.
\end{prop} 

The proof is by contradiction. Assume that the asymptotic motion is
bounded. Then all the $\mu_I$ must become vanishingly small due to the
explicit exponential $q^0$ dependence. Thus for sufficiently early time
(small volume), we can use the near form of potential valid for $\mu_I
\ll 1$.

Consider now just the radial motion, $R(q^0)$. Since it is bounded, it can 
either tend to a limiting value $R_*$ or it will keep oscillating. If it
has a limiting value, then we must have $\dot{R}, \ddot{R} = 0$ at $R =
R_*$. However, for the near form of the potential, its radial derivative
is {\em negative}. The radial equation of motion (\ref{Radial}) then implies
a contradiction unless possibly $R_* = 0$. 

For $R \sim 0$, we have,
\begin{equation} \label{NearOrigin}
W^{\text{near}}_{j,l}(R, \theta) \approx -2 (2j)^{-\Case{3 -l}{1 -l}} 
\left(\frac{3}{1 + l}\right)^{\Case{1}{1 - l}}
e^{\frac{2}{\sqrt{3}} (\frac{4 - 3l}{1 - l}) q^0} \left( 3 + \Case{8 - 12l +
6 l^2}{(1 - l)^2} R^2 \right)\ .
\end{equation}

Since at $R = 0$ the potential is negative and $h^{-1}$ must be strictly
positive, we must have $(\dot{R}^2 + R^2 \dot{\theta}^2) > 1/2$. But
this is impossible at $R_* = 0$. Thus, $R(q^0)$ {\em cannot have any finite
limiting value.}

If $R(q^0)$ oscillates, then it must have several extrema. Now the
radial equation (\ref{Radial}) together with the fact that
$\frac{\partial W^{\text{near}}_{j,l}}{\partial R} < 0$ implies that
$\ddot{R}$ must be {\em positive at any extremum}. Thus the radial
motion can have at the most one minimum and oscillatory motion is ruled
out. Since both the oscillatory and the limiting radial motions are
ruled out, we conclude that the trajectory cannot be confined to a
bounded region and the proof is completed.

For analyzing the case of unbounded motion, it is convenient to use a
vector notation:
\begin{eqnarray}\label{VecDefs}
\mu_I & := & \frac{1}{2j} e^{\rho_I q^0 + \vec{\sigma}_I \cdot \vec{q}}
~~,~~
\rho_I = \frac{2}{\sqrt{3}} ~~=: \rho_0 ~~~~ \forall ~~~ \text{I ~=~ 1, 2,
3} \nonumber \\
\vec{\sigma}_1 & := & \left( \frac{4}{\sqrt{6}} , 0 \right) ~,~
\vec{\sigma}_2 := \left( -\frac{2}{\sqrt{6}} , \sqrt{2} \right) ~,~
\vec{\sigma}_3 := \left( -\frac{2}{\sqrt{6}} , - \sqrt{2} \right)
\\
\vec{\sigma}_I^2 & = & \frac{8}{3} ~~,~~ 2 \rho_0^2 = \sigma_I^2 ~~,~~
\sum_I \vec{\sigma}_I = 0 \nonumber
\end{eqnarray}
In addition, we also have the expression for $\mu_I$ in terms of $R,
\theta$ given in equation (\ref{MuDefs}).

\begin{figure}[htb]
\begin{center}
\includegraphics[width=12cm]{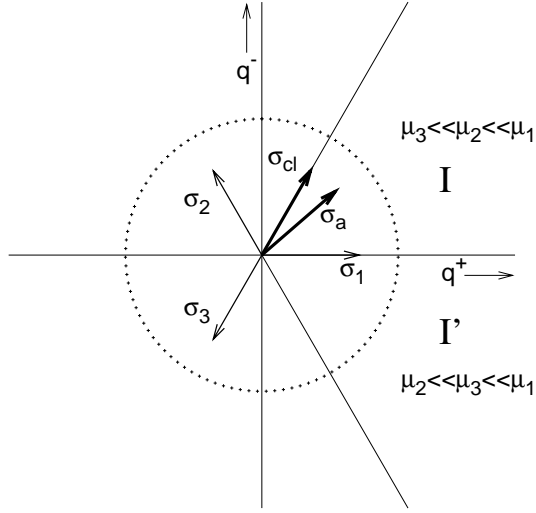}
\end{center}
\caption{The asymptotic regions I and I$^{\prime}$ of the anisotropy
plane. $R$ is outside the circle. $\vec{\sigma}_a$ varies with the
ambiguity parameter $l$ staying within I. The vectors shown are unit
vectors.
\label{Regions}}
\end{figure}

Observe that the ratios of the $\mu_I$'s do not have any explicit
dependence on $q^0$ and since $R(q^0)$ is very large, these ratios
either diverge, vanish or become equal to 1 (the last case only along
three special directions). Values of individual $\mu_I$'s however
depend on the relative growth of $R(q^0)$ and $|q^0|$. We will say:
$R(q^0)$ is {\em dominant} if $\Case{R(q^0)}{|q^0|} \to \infty$, {\em
sub-dominant} if $\Case{R(q^0)}{|q^0|} \to 0$ and {\em marginal} if
$\Case{R(q^0)}{|q^0|} \to \Case{\xi}{\sqrt{2}}$, where $\xi > 0$ is
constant such that $\xi = 1$ corresponds to a Kasner motion.  For
sub-dominant $R(q^0)$ and marginal one with $\xi \le 1$, all $\mu_I$'s
vanish (generically in the marginal case) while for dominant and $\xi
> 1$ cases, two of the $\mu_I$'s diverge (vanish) while the third one
vanishes (diverges) since the product of all $\mu_I$'s vanishes. Since
$2j > 1$, all $\mu_I$ are either much smaller or larger than 1 so that
the corresponding forms of the $F(\mu_I)$ can be used to approximate
the effective potential. To get explicit forms, we have to consider
various cases. Noting that the potential is invariant under cyclic
permutation of the $\mu_I$, it suffices to consider the cases: ($I$)
$\mu_3 \ll \mu_2 \ll \mu_1$ and ($I^{\prime}$) $\mu_2 \ll \mu_3 \ll
\mu_1$ . It is easy to see that ($I$) corresponds to large $R$ and
$\vec{\sigma_3}\cdot\vec{q} < \vec{\sigma_2}\cdot\vec{q} <
\vec{\sigma_1}\cdot\vec{q}$ from which it follows that $\vec{q}$ must
be confined to $0 < \theta < \Case{\pi}{3}$.  Likewise, the region
($I^{\prime}$) corresponds to large $R$ and $ -
\Case{\pi}{3} < \theta < 0$ (see figure \ref{Regions} ). 

Within these regions (ordering of the $\mu_I$'s), we have further sub-cases,
\begin{equation} \label{SubCases}
\text{(a)} ~  \mu_3 \ll \mu_2 \ll \mu_1 \ll 1 ~,~
\text{(b)} ~  \mu_3 \ll \mu_2 \ll 1 \ll \mu_1 ~,~ 
\text{(c)} ~  \mu_3 \ll 1 \ll \mu_2 \ll \mu_1 ~.
\end{equation}
The case where all $1 \ll \mu_I$, would imply that their product is
larger than 1 which is not possible for $q^0 \to - \infty$.

For these cases, we will derive the asymptotic form of the potential and
analyze the equations to deduce that only marginal $R(q^0)$ is possible
with $\xi = 1$. 

To derive the asymptotic potential, we prove a couple of lemmas:
\begin{lemma} \label{lem1}
 If $R(q^0) \to \infty$ as $q^0 \to - \infty$ then for any 
 real numbers $l, m, n$, $(\mu_I)^l (\mu_J)^m (\mu_K)^n$ either
 diverges, vanishes or goes to $(2j)^{-(l + m + n)}$.
\end{lemma}

The proof is obvious since $\mu_I \sim e^{\frac{2}{\sqrt{3}} q^0 R(q^0)
\cos \theta_I(q^0)}$ and asymptotically, the exponent is either 
positive, negative or zero. 

\begin{lemma} \label{lem2}
The spin connection components obey the inequalities:
\begin{center}
$
\begin{array}{llcl}
(a) & 1 \gg \mu_I \gg \mu_J \gg \mu_K & \Rightarrow & \Gamma_K \gg \Gamma_J 
\gg \Gamma_I ~ , ~ \Gamma_K \gg \Gamma_J \Gamma_I \ ; \\
(b) & \mu_I \gg 1 \gg \mu_J \gg \mu_K & \Rightarrow & \Gamma_K \gg \Gamma_J 
\ , \ \Gamma_K \gg \Gamma_I ~,~ \Gamma_K \gg \Gamma_J \Gamma_I \ ; \\
(c) & \mu_I \gg \mu_J \gg 1 \gg \mu_K & \Rightarrow & \Gamma_K \gg \Gamma_J 
\gg \Gamma_I ~ , ~ \Gamma_K \gg \Gamma_J \Gamma_I \ . 
\end{array}
$
\end{center}
\end{lemma}
To prove the above lemma, it should be noted that the ratio of any two
spin connections, say $\Gamma_I$ and $\Gamma_J$ can be written as
\begin{equation}
\frac{\Gamma_I}{\Gamma_J} = \left(\frac{\mu_J}{\mu_I}\right)
\left[ \frac{\frac{F(\mu_K)}{\mu_K} + \frac{F(\mu_J)}{\mu_J} -
{F(\mu_I)}^2}
{\frac{F(\mu_K)}{\mu_K} + \frac{F(\mu_I)}{\mu_I} -
{F(\mu_J)}^2}\right] ~,
\end{equation}
where the term inside the parenthesis is explicitly in reverse order.
Also it can be shown that for all cases and for all allowed values of
$l$, the term inside the bracket asymptotically {\em either} vanishes 
{\em or} goes to a constant {\em or} diverges but the divergence is
weaker than the convergence of $\left(\frac{\mu_J}{\mu_I}\right)$.

It follows that the potential takes the form $W_{j,l}(\vec{\mu}) \approx -8
j^2 \mu_I \mu_J \Gamma_K$. From Lemma \ref{lem1}, it follows that $\Gamma_K$
will become a monomial in the $\mu$'s. The potential then takes the
form:
\begin{eqnarray}
W_{j,l}^{(a)} (\vec{\mu}) &\simeq &-~ 4 j^2\left(\frac{3}{l+1}\right)^
{\frac{1}{1-l}} {\mu_I}^{\frac{2-l}{1-l}} {\mu_J}^2, \label{CASEA}\\
W_{j,l}^{(b)} (\vec{\mu}) &\simeq &-~ 4 j^2\left(\frac{3}{l+1}\right)^
{\frac{1}{1-l}} {\mu_I}^2 {\mu_J}^{\frac{2-l}{1-l}} \label{Wab}\nonumber \\
&\text{or}& - 4 j^2 {\mu_J}^2 \label{CASEB}\\
&\text{or}& - 4 j^2\left\{1 + \left(\Case{3}{1 + l}\right)^
{\frac{1}{1 - l}} (2 j)^{-\frac{2-l}{1-l}}\right\}
{\mu_J}^2 , \nonumber \\
W_{j,l}^{(c)} (\vec{\mu}) &\simeq &  -~ 4 j^2 {\mu_I}^2 ~. \label{CASEC}
\end{eqnarray}
Thus we have proved,
\begin{prop} \label{PropAsy}
In all cases where the $\mu_I$ either vanish or diverge, the potential 
necessarily takes the form,
\begin{equation}\label{AsyForm}
W_{j,l}(R, \theta, q^0) ~ \sim ~ - C e^{\rho q^0 + \vec{\sigma}\cdot\vec{q}} 
\end{equation}
where $C, \rho, \vec{\sigma}$ are constants with $C > 0$, $\rho > 0$
and $\vec{\sigma}^2 \leq 2\rho^2$. 
\end{prop}

Thus in all the cases, the potential has the form (\ref{AsyForm}). In fact, 
for the sub-case (a) and the first of the sub-case (b) above, we have 
$\vec{\sigma}^2/(2\rho^2)=(3l^2-6l+4)/ (3l-4)^2$, which is smaller than one 
for all $0<l<1$ (and exactly one for the limiting case $l=1$). In the
remaining two of the sub-case (b) and the sub-case (c), we have 
$\vec{\sigma}^2/ (2\rho^2)=1$. Notice that the asymptotic potential is 
{\em negative}. Together with $h > 0$, this implies that $1 - 2 
\dot{\vec{q}}^2 < 0$.

Using the asymptotic form of the potential, the Cartesian form 
(\ref{Cartesian}) of the equations becomes ($\vec{v} := \dot{\vec{q}}$), 
\begin{equation}
\dot{\vec{v}} ~=~ \left( \vec{v}^2  - \frac{1}{2} \right)
\left( \frac{\vec{\sigma}}{2} + \rho \vec{v} \right) ~~,~~
\vec{v}^2 > \frac{1}{2} ~~,~~ 2 \rho^2 - \vec{\sigma}^2 \ge 0 \ .
\label{VelEqn}
\end{equation}

Notice that the above equation is a {\em first order, autonomous}
equation for the velocity. The (vector) equation has fixed points at
$\vec{v}^2 = \Case{1}{2}$ and at $\vec{v} = - \Case{\vec{\sigma}}{2
\rho}$. For the latter, isolated fixed point, $\vec{v}^2 = \Case{1}{2}
\Case{\sigma^2}{2 \rho^2} \le \Case{1}{2}$ and hence this is {\em not}
accessible. 

Taking scalar product of the equation (\ref{VelEqn}) with $2 \vec{v}$ one
gets,
\begin{equation} \label{DotEqn}
\dot{\vec{v}^2} = \left(\vec{v}^2 - \frac{1}{2}\right) (2 \rho
\vec{v}^2) \left(
\frac{\vec{\sigma}\cdot\vec{v}}{2 \rho \vec{v}^2} + 1\right) \ .
\end{equation}

The inequalities on the $\vec{v}^2$ and $\vec{\sigma}^2, \rho^2$ imply
that the third factor is always {\em positive} and hence $\vec{v}^2$ must 
{\em decrease} as $q^0$ is {\em decreased}. Clearly, $\vec{v}^2$ is
bounded {\em above} by $(\vec{v}_{\text{initial}})^2$ and bounded {\em
below} by $\Case{1}{2}$. Since the equation for $\vec{v}^2$ is
monotonic, asymptotically $\vec{v}^2$ must reach the lower bound.

Taking the `cross product' of the equation (\ref{VelEqn}) with
$\vec{\sigma}$ one obtains,
\begin{equation}\label{CrossEqn}
\dot{(\vec{\sigma} \times \vec{v})} = \rho \left(\vec{v}^2 - 
\frac{1}{2}\right) 
(\vec{\sigma} \times \vec{v}) 
\end{equation}

This equation shows that either $\dot{\vec{\sigma} \times \vec{v}} = 0$
or, if non-zero, cannot change sign. If $\phi$ denotes the angle
between $\vec{v}$ and $\vec{\sigma}$, then the cross product is $v
\, \sigma \, \text{sin}\phi$ ($v := |\vec{v}|, \sigma := |\vec{\sigma}|$) 
which is clearly bounded. Hence it must also approach a limiting value. 
Since both the magnitude of the velocity as well as its component orthogonal
to $\vec{\sigma}$ approach a limiting value, we conclude that asymptotically
$\vec{v}$ must approach {\em some} $\vec{v}_*$ such that $\vec{v}_*^2 = 
\Case{1}{2}$. This also means that {\em $R(q^0)$ must be marginal with 
$\xi = 1$}, both the dominant and the sub-dominant as well as marginal
with $\xi \ne 1$ possibilities for $R(q^0)$ being ruled out. Since in
the sub-cases listed in equation (\ref{SubCases}), (b) and (c) have at
least one of $\mu_1, \mu_2$ to be larger than 1 and for $\xi = 1$
marginal behavior all $\mu_I$'s must be less than 1, these two
sub-cases are ruled out. Thus, sub-case (a) is the only relevant case
and, corresponding to region (I), we have $\vec{\sigma}
=\vec{\sigma}_a$ which is read off from equation (\ref{CASEA}) as,
\begin{equation}\label{SigmaADef}
\vec{\sigma}_a ~=~ \frac{2 -l}{1 -l} \vec{\sigma}_1 + 2 \vec{\sigma}_2 \ .
\end{equation}
As shown in the figure \ref{Regions}, $\vec{\sigma}_a$ lies within the 
region (I).

To study the nature of a candidate fixed point $\vec{v}_*$ of equation
(\ref{VelEqn}), linearize the equation by putting $\vec{v} = \vec{v}_*
+ \delta \vec{v}$. This leads to,
\begin{equation} \label{LinEqn}
 \dot{\vec{\delta v}} ~=~ (\vec{v}_* \cdot \vec{\delta v }) \vec{\Delta}
~~~,~~~ \vec{\Delta} :=  \vec{\sigma} + 2 \rho \vec{v}_* \ .
\end{equation}

Clearly, $\vec{\delta v}$ orthogonal to $\vec{v}_*$ is a zero mode of
the linearized equation which does not evolve. Taking scalar product
with $\vec{v}_*$ gives, using $\vec{v}_*^2 =\frac{1}{2}$,
\begin{equation}
\dot{(\vec{v}_* \cdot \vec{\delta v})} = (\vec{v}_* \cdot \vec{\delta v
}) ( \vec{v}_* \cdot \vec{\sigma} + \rho)\,.
\end{equation}

Linear stability of $\vec{v}_*$ thus demands $\rho + \sigma (\cos
\phi_*)/\sqrt{2} > 0$. Since $\sqrt{2}\rho/\sigma \ge 1$, this
condition is satisfied {\em for all $\phi_*$}. Thus we have proved
that {\em the asymptotic velocity under the evolution equation
$(\ref{VelEqn})$ is a Kasner velocity and every Kasner velocity is
admissible.} Furthermore, this velocity is approached {\em
exponentially} as implied by the stability analysis.

One can also derive a relation between an initial velocity and the
limiting velocity. Denote: $v_{\parallel} := v\cos \phi$,
$v_{\perp} := v\sin \phi$. Taking scalar and cross product of equation
(\ref{VelEqn}) with $\vec{\sigma}$ gives,
\begin{equation} \label{AlphaBetaEqn}
\dot{v}_{\parallel} ~ = ~ \left( \vec{v}^2 - \frac{1}{2} \right) \left(
\frac{\sigma}{2} + \rho v_{\parallel}\right) ~~,~~
\dot{v}_{\perp} ~ = ~ \left( \vec{v}^2 - \frac{1}{2} \right) \left(
\rho v_{\perp}\right) ~~.
\end{equation}

If $v_{\perp} = 0$ initially, then it is identically zero and
therefore $\phi_* = 0 \ \text{or} \ \pi$ depending upon the sign of
$v_{\parallel}$. The $v_{\parallel}$ equation can be easily integrated
but we do not need the explicit solution. If $v_{\perp} \ne 0$
initially then it must retain its sign. In this case taking the ratio
$\dot{v}_{\parallel}/\dot{v}_{\perp}$ gives a differential equation
for $v_{\parallel}$ as a function of $v_{\perp}$ which can be
immediately integrated to give an {\em orbit equation},
\begin{equation} \label{OrbitEqn}
v_{\perp} ~=~ A \left( v_{\parallel} + \frac{\sigma}{2 \rho} \right)
\end{equation}
where $A$ is a constant of integration.

The orbit is a straight line in the velocity plane always passing
through the point $v_{\parallel} = - \Case{\sigma}{2 \rho} \ (\, > - \Case{1}
{\sqrt{2}}\, )\, , v_{\perp} = 0 $ and with a slope determined 
by $A$. This point is exactly the isolated fixed point of the (\ref{VelEqn}) 
equation.

Note that since $v_{\perp} \ne 0$ neither $A$ nor ($v_{\parallel} + 
\Case{\sigma}{2 \rho}$) can be zero for any finite $q^0$. We also know 
that asymptotically $v_{\perp} \to v_{\perp *} := \Case{1}{\sqrt{2}} 
\sin \phi_*$ and $v_{\parallel} \to v_{\parallel *} := \Case{1}
{\sqrt{2}} \cos \phi_*$. As noted earlier, only $\vec{\sigma}_a$ 
is consistent with asymptotic Kasner motion and for this case $2 \rho_a^2 
- \sigma_a^2$ is strictly positive. This implies that $(v_{\parallel *} + 
\Case{\sigma}{2 \rho})$ is non-zero and therefore $v_{\perp *}$ is also 
non-zero. Evaluating equation (\ref{OrbitEqn}) for initial values of 
($v_{\parallel}, v_{\perp}$) determines $A$ as a function of $v, \phi$ 
while evaluating the same equation for asymptotic values ($v_{\parallel *}, 
v_{\perp *}$) determines the angle $\phi_*$. 
Explicitly, in terms of $A$,
\begin{eqnarray} \label{AngleEqn}
\cos\phi_* &=& \frac{1}{1 + A^2}\left\{ - A^2
\frac{\sigma}{\sqrt{2}\rho} ~ \pm ~ \sqrt{1 + A^2 - A^2 \frac{\sigma^2}{2
\rho^2} } ~ \right\} \nonumber \\
\sin\phi_* &=& \frac{A}{1 + A^2}\left\{ 
\frac{\sigma}{\sqrt{2}\rho} ~ \pm ~ \sqrt{1 + A^2 - A^2 \frac{\sigma^2}{2
\rho^2} } ~ \right\} 
\end{eqnarray}

The two roots ($\pm$ signs) for sin($\phi_*$) always have opposite signs
and since the sin($\phi$) does not change sign, the initial condition
sets the sign and selects a unique limiting value $\phi_*$.

Now the trajectory in the anisotropy plane can be obtained trivially as,
$\vec{q}(q^0) = \vec{v}_* q^0 + \hat{q}$ which leads to,
\begin{equation} \label{Soln}
R(q^0) ~ \sim ~ - \frac{1}{\sqrt{2}} q^0 + O(1) ~~~~,~~~~
\theta(q^0) ~ \sim ~ \theta_0 + O\left(\frac{1}{\sqrt{- q^0}}\right) \ ,
\end{equation}
where $\theta_0 = \phi_*$ plus the angle $\vec{\sigma}$ makes with the
$q^+$ axis. Notice that asymptotically $\vec{q}$ points in the direction
{\em opposite} to that of $\vec{v}_*$.

Let us now compare a similar analysis for the classical potential. In
the classical case, one does not need to distinguish the sub-cases of
(\ref{SubCases}) as the potential is directly given as a sum of monomials
in $\mu_I$. The asymptotic form in region (I) is the same as given in
Proposition \ref{PropAsy} with $C < 0, \rho = \Case{4}{\sqrt{3}}, 
\vec{\sigma} = - 4 \vec{\sigma_3} = ( \Case{8}{\sqrt{6}}, 4 \sqrt{2} )$. 
This implies two differences compared to the effective case:
$\Case{\sqrt{2}\rho}{\sigma} = \Case{1}{2} < 1$ and $\vec{v}^2 <
\Case{1}{2}$. The first order, autonomous equation for the velocity
remains the same as (\ref{VelEqn}) (except for the inequalities) and
so do the equations (\ref{DotEqn}, \ref{CrossEqn}).  The isolated
fixed point of (\ref{VelEqn}) gets ruled out as before.  Unlike the
effective potential case, the third factor in (\ref{DotEqn}) no
longer has a fixed sign. If it is zero, then $v^2$ gets frozen at a
fixed value and if it is non-zero, it cannot change its sign. Since
the evolution of $v^2$ is bounded, once again we get that $v^2 \to
v_*^2$ {\em although we cannot conclude at this stage that $v^2_* =
\Case{1}{2}$}. As before, $v \sin \phi$ is bounded and evolves
monotonically thereby reaching some limiting value. Thus we conclude
that $\vec{v} \to \vec{v}_*$ as before though without the condition $v^2_* =
\Case{1}{2}$.

Properties of a candidate limiting velocity are analyzed by
linearization, $\vec{v} = \vec{v}_* + \delta \vec{v}$. This leads to,
\begin{equation} \label{ClassLinEqn}
\dot{\vec{\delta v}} ~=~ \gamma \vec{\Delta} + \gamma \rho \delta
\vec{v} + (\vec{v}_* \cdot \vec{\delta v }) \vec{\Delta}
~~~,~~~ \vec{\Delta} :=  \vec{\sigma} + 2 \rho \vec{v}_* ~~~,~~~ \gamma :=
v_*^2 - \frac{1}{2} \ .
\end{equation}

Clearly, $\delta \vec{v} \to 0$ must hold and therefore either $\gamma =
0$ or $\vec{\Delta} = \vec{0}$. $\vec{\Delta} = \vec{0}$ is ruled out
since this gives $v^2_* = 2$ and hence $\gamma = 0$ must hold. We have
recovered the $v_*^2 = \Case{1}{2}$. The rest of the stability analysis
now gives, exactly as before $(\rho + \Case{\sigma \, \text{cos}\, 
\phi_*}{\sqrt{2}}) > 0$ or cos($\phi_*$) $> - \Case{1}{2}$. Unlike the
effective potential case, now we have a restriction on the asymptotic
angles $\phi_*$. The angle $\phi_*$ can again be obtained as in eq.
(\ref{AngleEqn}), however due to $\Case{\sigma^2}{2 \rho^2} = 4$,
the square roots are real only if $A^2 \le \Case{1}{3}$.

\begin{figure}[htb]
\begin{center}
\begin{tabular}{cc}
\includegraphics[width=8cm]{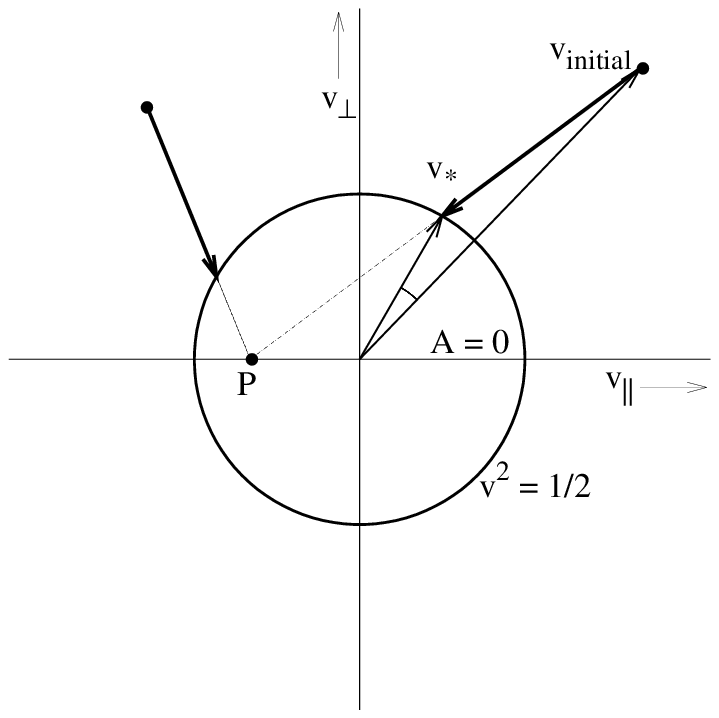} &
\includegraphics[width=8cm]{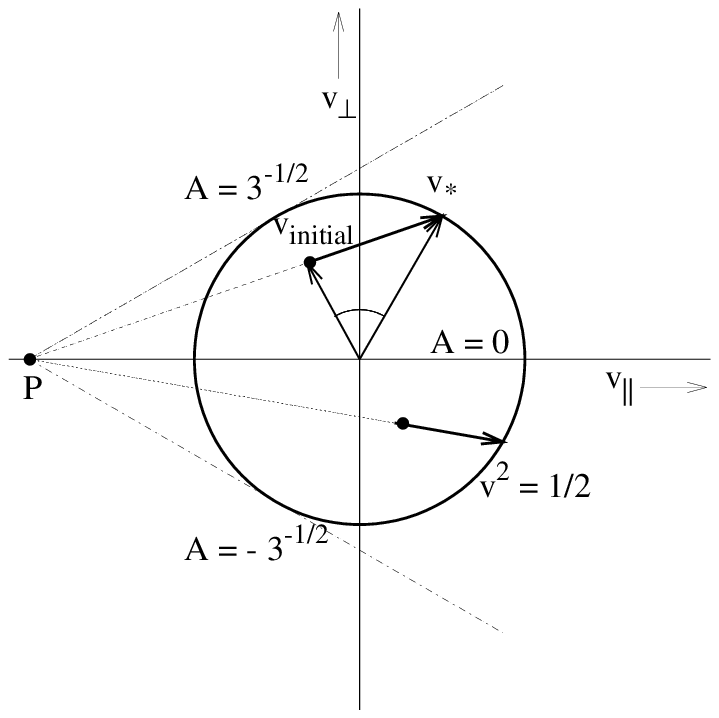} \\
(a) \hspace{0.7cm} & (b) \hspace{0.7cm}
\end{tabular}
\end{center}
\caption{ The point $P$ and the circle constitute the fixed points. The
heavy, directed straight line segments are typical orbits of the velocity
dynamics in the velocity plane. The $v_{\parallel}, v_{\perp}$ are relative
to $\vec{\sigma}$. The angle shown is the total change in the direction of
the velocity vector. The figure (a) corresponds to the effective potential 
while (b) corresponds to the classical potential.
\label{Orbits}}
\end{figure}

The orbit equation provided a very simple pictorial representation
shown in Figure \ref{Orbits}. The two differences between the
effective potential and the classical potential control the {\em
location} of the isolated fixed point ($v_{\perp}=0$) and the setting
of initial conditions. For both potentials the continuum of fixed
points is determined by the circle: $v^2_* = \Case{1}{2}$. For the
effective potential, the isolated fixed point lies inside the circle
while the initial condition must be set outside the circle. For the
classical case it is exactly the opposite. The monotonic decrease of
$v^2$ in the case of the effective potential and monotonic increase of
$v_{\parallel}$ in the case of the classical potential shows the
direction of traversal of the orbits. The exponential approach to
$\vec{v}_*$ follows from the linear stability analysis.

How does one understand the asymptotic Kasner solutions for the classical 
potential given the well known oscillatory behavior? Recall that the
asymptotic direction reached in the anisotropy plane is {\em opposite}
to that of $\vec{v}_*$ (since $q^0$ is negative). The stable Kasner
velocity directions are confined to $- \Case{2 \pi}{3}$ to $ \Case{2
\pi}{3}$ relative to $\vec{\sigma}$ which makes an angle $\Case{\pi}{3}$
to the $q^+$ axis. The corresponding trajectories therefore satisfy
$\Case{2\pi}{3} < \theta_0 < 2 \pi$ which {\em precisely excludes} the
region (I)! The qualitative analysis rests on the assumption that
(\ref{VelEqn}) remains valid throughout the evolution. This assumption
must fail in the case of the classical potential --- a trajectory that
enters an asymptotic region (I, I$^{\prime}$ or their $120^0$
rotations), must exit it in a finite $q^0$ interval. By contrast, for the
effective potential, since {\em all} $\phi_*$ are stable, there always
exist directions such that $\theta_0$ is in the asymptotic region. The
remaining directions will again correspond to exiting the region only to
eventually enter another region with compatible $\theta^0$.

Combining similar analyses of all other asymptotic regions we conclude
that {\em any trajectory of the effective dynamics will always go
exponentially to a Kasner trajectory as volume goes to zero, with at the 
most finitely many inter-region traversals (oscillations).}

\section{Summary and Discussion}

Bianchi IX models are well known to display the most complex behavior
close to a {\em homogeneous} singularity but in addition, in the BKL
scenario, play a pivotal role in suggesting a strongly spatially
dependent structure of the spatial slice close to a {\em generic,
inhomogeneous} cosmological singularity. The primary source of this
picture is the chaotic behavior of the Bianchi IX dynamics which
forces infinite fragmentation with no correlations among the
fragments. Since chaotic behavior itself is a consequence of
unbounded growth of the curvature invariants, i.e.\ the singularity,
and since the loop quantization of diagonal, homogeneous models has
been shown to be singularity free, it is natural to focus on the
modification due to the loop quantization of the vacuum Bianchi IX model.

Since the chaotic dynamics is best understood in the classical
setting, we worked with an effective classical dynamics from the
underlying, singularity free quantum dynamics. This was done by
exploiting the (always) available continuum approximation together
with a WKB ansatz for the continuum wave function. The most
significant, qualitatively distinct effects of the loop quantization
got incorporated into the effective potential via the definition of
inverse triad components together with the quantization ambiguity
parameters ($j \ge \Case{1}{2}, 0 < l < 1$). The effective potential
has several qualitative differences from the classical potential. (i)
Its volume dependence is not in a factorizable form which makes the
dynamical equations explicitly `time' ($q^0$) dependent. (ii) It does
have walls for large volume but unlike the classical potential these
walls are not confining -- they have a finite height and a finite
extent. (iii) For smaller volume, these walls just disappear, the
potential becomes negative everywhere and approaches zero in the zero
volume limit. Qualitatively one expects a simpler, non-oscillatory,
behavior asymptotically resembling Kasner behavior since the
potential vanishes.

We presented the details of the effective dynamics at two levels. In
section \ref{WithinDomain}, we stayed entirely within the domain of
validity of the effective model and demonstrated the {\em
non-asymptotic} stability of the Kasner motion to the inclusion of the
effective potential. The vanishing of inverse triad eigenvalues for
small triad eigenvalues was directly responsible for this stability. In
section \ref{Extrapolation}, we extrapolated the effective
dynamics beyond its natural domain of validity into the small volume
regime, much as one does in the usual classical analysis. We proved in
detail that for every initial condition, after at most finitely many
oscillations, the trajectory will diverge to infinite anisotropies as a
Kasner motion. The classical case was also contrasted. All the proofs
work uniformly for all the allowed values of the ambiguity parameters
demonstrating the robustness of loop quantum effects.

To explore the possible implications of the results for the structure of
inhomogeneous singularities, let us assume the validity of the BKL picture
and divide up a spatial slice into `cells' which are approximately
homogeneous and have (initial) anisotropies parameterized by Bianchi IX.
We may picture the so decomposed initial spatial slice as a `sprinkling
of points' in the Bianchi IX anisotropy plane, each point corresponding
to a cell. The approach to an inhomogeneous singularity can now be viewed
as evolution of the initial points under a Bianchi IX effective (or
otherwise) dynamics. 

When the evolution is chaotic (as in the classical case), two nearby points
will diverge away, effectively loosing initial correlations implied by the 
smooth geometry of the slice. This needs further decomposition and so on. 
Asymptotically then there would be infinitely many points performing {\em 
un-correlated} motion in the anisotropy plane -- i.e. a non-interacting 
two dimensional `gas'.

When the chaotic motion is replaced by asymptotic Kasner evolution (as
with the effective dynamics), the points will move in a simpler manner
getting distributed into the various regions and after some initial
crossings between regions they will remain confined to their respective
regions. Their correlations, just before this stage, will not be washed
away and we will have an `interacting gas.' In both cases, all
angular sectors are `filled' suggesting that a sufficiently large
class of inhomogeneities can be parameterized as a distribution of
anisotropies to the cells. (The inhomogeneities would be arbitrary in the
case of chaotic dynamics.) 

Just for contrast assume that Bianchi IX dynamics is governed by the
negative of the classical potential. Since the walls disappear, this
is a trivial way of removing chaos (by hand). The stable fixed points
of the asymptotic dynamics now include all those for the classical
potential but in addition the isolated fixed point at $\sigma_{\perp}
= 0, \sigma_{\parallel} = - \Case{\sigma}{2 \rho} = - \sqrt{2}$ is
also accessible. Now the asymptotic motion {\em compatible with the
region} would be to the isolated fixed point. By permutation, one will
have just three possible asymptotic velocities (along the dips in the
classical potential). All the initially distributed points will
cluster in these directions and the singularity will look essentially
homogeneous.

We will now use these three scenarios, classical potential, negative
classical potential, and effective potential, to shed light on the
general structure of classical singularities. In the context of the
approach to a classical singularity it has been observed
\cite{PenroseSing} that the beginning and the end of a universe like
ours appear very different from each other. Our observable part of the
universe starts out very homogeneous and isotropic, but turns much
more complicated in the future. Black holes form and their
singularities make space more and more inhomogeneous. Let us now
assume that the universe may re-collapse. Then by time reversal
symmetry the structure of both the initial and the final singularity
at the boundaries of space-time would have to be the same. If the
universe does form black holes (which once formed will not disappear
in a classical context), then the inhomogeneity implied by their
presence will not disappear implying a strongly inhomogeneous
structure for the final singularity.  From what we can tell by
observations of our own universe, then, the situation looks far
from time reversal symmetric. Right now space is very homogeneous and
isotropic at large scales, and is expected to have been even more
homogeneous earlier, while we do not see a reason for it to homogenize
away the structure implied by black holes in the future.

The large degree of homogeneity is usually explained by referring to
inflation, which expands a small patch of an initial space and thus
removes inhomogeneities. (However, whether inflation begins at all in
an inhomogeneous context is apparently an unresolved issue
\cite{GFREllis,Raychaudhuri}. Moreover, if the classical BKL dynamics
is true, this initial space close to the classical singularity would
have structure at arbitrarily small scales which would be difficult to
smooth out even for inflation. If the chaotic behavior leads to a
fractal initial state, inhomogeneities would not be removed at all due
to self-similarity.) If the inhomogeneous structure is determined by
the chaotic Bianchi IX dynamics, then the primordial anisotropies
would be completely arbitrary in such a patch. However, there are
arguments to the effect that any such anisotropies would be washed
away before inflation ends \cite{Sahni}.  This would be good for
phenomenology in that one does not have to worry about the details
prior to the beginning of inflation but then one has also lost any
contact with whatever quantum effects that would be present in the
earlier epoch. Alternatively, those effects could be influential in
determining the beginning of inflation itself. Deciding this may be
harder when homogeneous patches are {\em un-correlated}.  In any case,
the classical expectation for the structure of singularities would be
consistent with the expected inhomogeneous end, but could be
problematic for the beginning.

The hypothetical case of a negative classical potential considered
above implies homogeneous structure for the singularity. This would be
quite desirable for the initial singularity since it is compatible with
the observed large scale homogeneity. One may at the most have to
worry about the isotropization issue. However, it would be in conflict
with an inhomogeneous final singularity and hence inconsistent with
black hole formation during the evolution of the universe. Thus merely
removing the chaotic Bianchi IX dynamics would not be satisfactory.

The effective dynamics, finally, offers a third possibility. It admits
a sufficiently general class of inhomogeneities (thereby accommodating
black hole formation) but does not give rise to structure on
arbitrarily small scales and retains some correlations among the
patches. This may make starting inflation easier. Alternatively the
mechanism of inflation could be very different which will not only
homogenize but may also retain some imprint of the underlying quantum
dynamics. The consequences are not worked out at present but it does
open up other possibilities to be explored. In this aspect, matter
will almost certainly play a significant role. The influence of matter
on the effective dynamics as well as the other issues mentioned above
will be dealt with elsewhere.

While the effective dynamics is consistent with our expectations for
both the beginning and the end of the universe, the apparent time
reversal asymmetry remains. This is explained by the fact that the
situation is, in fact, time asymmetric due to our own position in the
universe. We can see only some part of it, not the whole space-time,
and in particular we see only a small part of the beginning. With the
current understanding, the observable part of our universe can well be
part of a classical space-time with a very inhomogeneous initial
singularity. Since most of the initial singularity is unobservable,
however, it is not discussed further. The final singularity, on the
other hand, is completely unobservable until it is reached. If we
compare only observable properties from within the universe, we simply
cannot possibly know enough to tell whether past and future
singularities are similar. If we compare the theoretical structure of
a space-time from outside, then we conclude that in fact there is no
conceptual difference between the beginning and the end of a generic
space-time. Only if we compare the observable part of the initial
singularity with the theoretical expectation for a final singularity
does the time asymmetry appear.

\begin{acknowledgments}
 We thank A.\ Coley, R.\ Penrose, A.\ Rendall and R.\ Vaas for discussions.
\end{acknowledgments}

\end{document}